%% file: TCMT_paper_JINST.tex
\pdfoutput=1
\documentclass{JINST}

\usepackage{epsfig}
\usepackage{rotate}  
\usepackage{cite} 
\usepackage{amssymb}
\usepackage{amsbsy}

\usepackage{amssymb,amsmath,array}
\usepackage{footmisc}

\title{Construction and performance of a silicon photomultiplier/extruded scintillator tail-catcher and  muon-tracker}

\input{calice_authors_tcmt_04jan2012.tex}

\abstract{
A prototype module for an International Linear Collider (ILC) detector was built, installed, and tested between 2006 and 2009 at CERN and Fermilab as part of the CALICE test beam program, in order to study the possibilities of extending energy sampling behind a hadronic calorimeter and to study the possibilities of providing muon tracking.
The "tail catcher/muon tracker" (TCMT) is composed of 320 extruded scintillator strips (dimensions 1000 x 50 x 5 mm$^{3}$) packaged in 16 one-meter square planes interleaved between steel plates.  The scintillator strips were read out with wavelength shifting fibers and silicon photomultipliers. The planes were arranged with alternating horizontal and vertical strip orientations.  Data were collected for muons and pions in the energy range 6~GeV to 80~GeV.  Utilizing data taken in 2006, this paper describes the design and construction of the TCMT, performance characteristics, and a beam-based evaluation of the ability of the TCMT to improve hadronic energy resolution in a prototype ILC detector.  For a typical configuration of an ILC detector with a coil situated outside a calorimeter system with a thickness of 5.5 nuclear interaction lengths,  a TCMT would improve relative energy resolution by 6-16\% for pions between 20 and 80 GeV. }


\begin{document}

\section{Introduction}
\label{sec:Section1}
Capturing the full physics potential of a next generation lepton collider such as the International Linear Collider (ILC) will require jet energy resolution nearly a factor of two beyond that of currently operating detectors. The improved jet resolution, approximately 30\%/$\sqrt{E/\mathrm{GeV}}$ at the Z mass and 3-4\% at all energies, is essential for complete and precise characterization of the Higgs boson and top quark as well as for sensitive searches for new physics \cite{LOIs}. In particle flow algorithms, which offer a promising avenue for improved jet resolution, tracking systems are used to reconstruct the momentum of charged particles while electromagnetic and hadronic calorimeters provide information on neutral electromagnetic and neutral hadronic particles \cite{PFA}. The reconstruction of individual showers in a jet requires fine transverse and longitudinal segmentation of the calorimeters in addition to good intrinsic energy resolution.

The CAlorimeter for LInear Collider Experiment (CALICE) Collaboration \cite{CALICE} has been evaluating prototype 
calorimeters and technologies which have the potential to achieve energy and spatial resolution required for next generation linear colliders. The prototype calorimeters also offer the opportunity to improve and benchmark Monte Carlo simulations, particularly important for describing hadronic showers.  Accordingly, the CALICE Collaboration has undertaken a program to illuminate complete (electromagnetic and hadronic) prototype calorimeter systems with electron, muon and hadron test beams.  From 2006 to 2009, beam tests were conducted at Deutsches Elektronen-Synchrotron (DESY), the European Organization for Nuclear Research (CERN), and Fermi National Accelerator Laboratory (FNAL).  The calorimeter system was composed of a sampling electromagnetic calorimeter  (either silicon-tungsten \cite{ECAL} or scintillator-tungsten depending upon the data run), followed by a steel and scintillator tile hadron calorimeter \cite{AHCAL}, and ending with the tail-catcher/muon tracker (TCMT) constructed of steel and extruded scintillator strips.  No magnet was incorporated into the test beam calorimeters.

In this paper, we describe the design, construction, and first beam tests of the TCMT.  The electromagnetic and hadronic calorimeters are described in detail elsewhere \cite{ECAL}, \cite{AHCAL}.  The TCMT, a sampling calorimeter, is composed of alternating layers of extruded scintillator and steel plates.  Light from the scintillator is read out with wavelength shifting fibers and silicon photomultipliers (SiPM) \cite{SiPM} 
which offer an opportunity to evaluate the performance of scintillator/SiPM calorimetry. In Section 2 we describe the TCMT in more detail. The TCMT increases the depth of the CALICE prototype which is important to explore the ultimate resolution of the calorimeter, the leakage associated with thinner, practical calorimeter systems, and the complete structure of hadronic showers.  The extended depth of the calorimeter also permits beam-based mock-ups to emulate and explore the performance of realistic detector systems including calorimeters, magnet coils, and post-coil sampling.  In addition,  a TCMT serves a dual function by offering an opportunity to increase the muon tracking capability of a detector system.  

Section 3 summarizes the CALICE installation at the CERN beamline and the data set collected in 2006. In Section 4 we offer information on the performance of the TCMT and Section 5 explores the response of the TCMT to hadronic showers.  In Section 6 we use the TCMT to evaluate the impact of the material of a magnet coil on an ILC calorimeter.  Finally, Section 7 provides a summary and conclusions.

\section{Design and Construction of the TCMT}
\label{sec:Section2}
A TCMT deepens calorimetric sampling beyond electromagnetic and hadronic calorimetry, which have typical depths of five to six nuclear interaction lengths, while simultaneously providing an opportunity to track and detect muons.  Thus, a TCMT must be of sufficient depth to contain the tail of hadronic showers and have sufficiently granular segmentation to serve as a muon tracker. 
A prototype TCMT should also be deep enough to offer the opportunity to simulate the impact of a superconducting magnet coil and the associated cryostat, mitigated by post-coil calorimetry.  Calorimeter design simulations for the proposed ILC Silicon Detector (or SiD) \cite{SID} were used to select design parameters including the thickness and width of the strips, number of active layers, and thickness of the absorber layers.

\begin{figure}[ht]
\begin{center}$
\begin{array}{c}
\includegraphics[width=110mm]{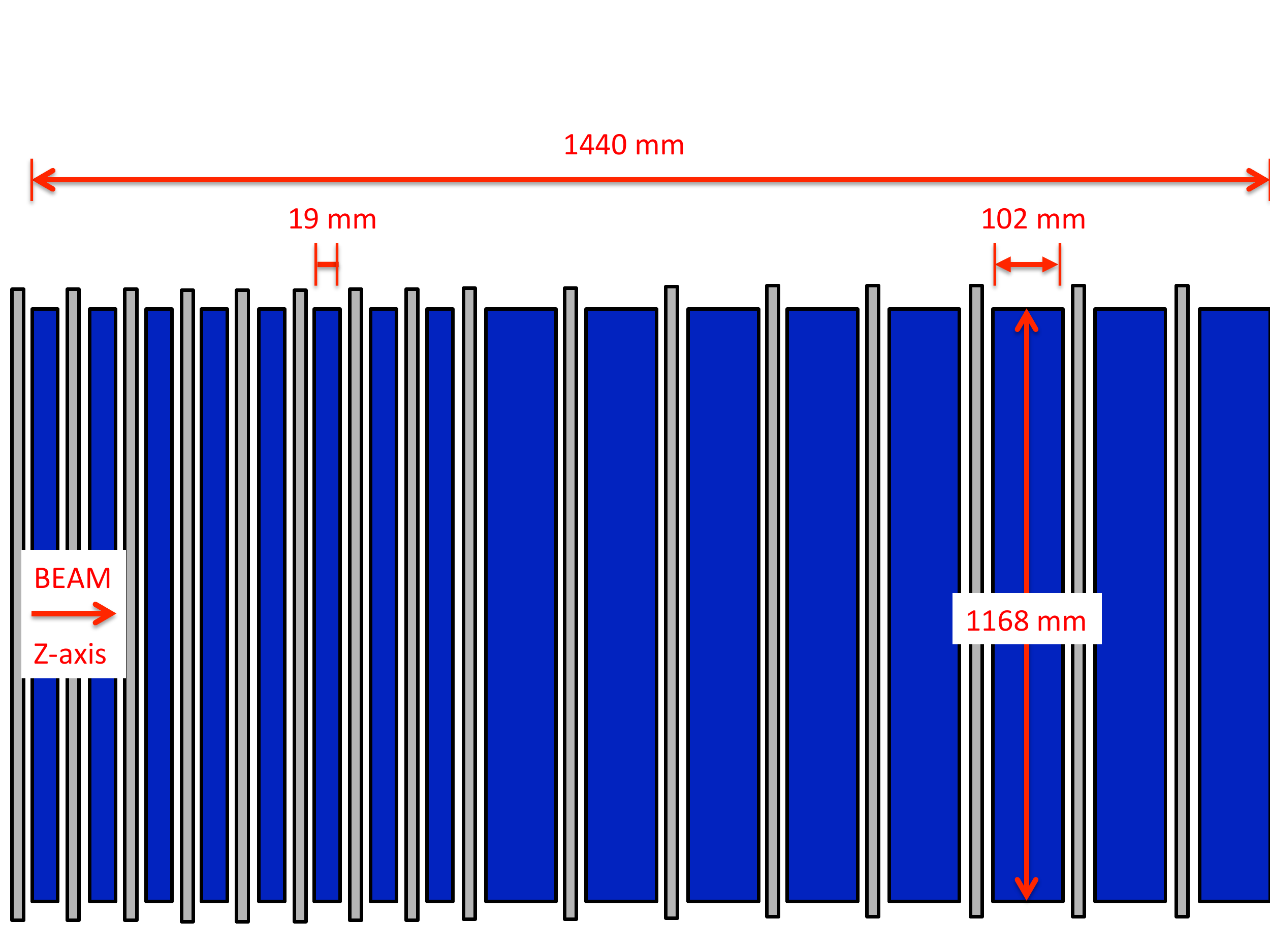} 
\vspace{-1.0cm} \\
\includegraphics[width=140mm]{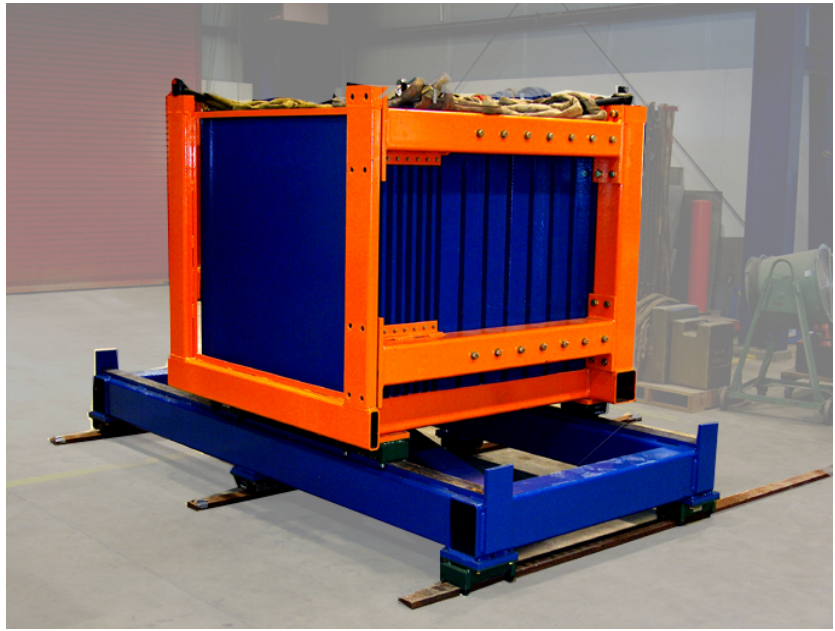}
\vspace{-1.0cm} \\
\end{array}$
\end{center}
\caption{Top:  Schematic (not to scale) showing alternating absorber plates (blue) and active medium (grey).  Bottom: TCMT steel frame showing the forward fine section (on the left) and the back coarse section.}
\label{fig:LayerSchematic}
\end{figure}

The TCMT is shown schematically and pictorially in Fig. \ref{fig:LayerSchematic}.  The TCMT mechanical structure, including the supporting frame, has a length of 1590 mm (along the beam direction), a height of 1090 mm, and a weight of approximately 10 tons.   As the figure indicates, the TCMT has a fine section with eight layers of 19 mm thick steel absorber plates  and a coarse section with  eight layers of 102 mm thick steel absorber plates. The fine section has a  thickness (or sampling fraction) close to that of the CALICE hadronic calorimeter permitting a uniform examination of the final portion of hadronic showers.  The absorber is composed of ASTM A36 steel (Fe 99\%, C 0.26\%, Mn 0.75\%, Cu 0.2\%, P 0.04\%, and S $<$0.05\%).   We take the absorption length to be that of pure iron or 
\mbox{131.9 $\textrm{g/cm}^2$}
(16.75 cm).   All plates have transverse dimensions of 1168 x 1168 mm$^2$ and are separated by 31.5 mm gaps. The structure accommodates 16 active planes. 

The total length of the sampling structure from the front of the first fine plate to the end of the last coarse plate is 1440 mm.  The sampled absorber (the last plate is not included) corresponds to 866 mm of steel or  a nuclear interaction length of  5.2 $\lambda_n$  which is equivalent to a pion interaction length of 4.2 $\lambda_{\pi}$. The mechanical structure and absorber stack were engineered and assembled by the Particle Physics Division at Fermi National Accelerator Laboratory.

\begin{figure}[tbh]
\centering
\includegraphics[width=150mm]{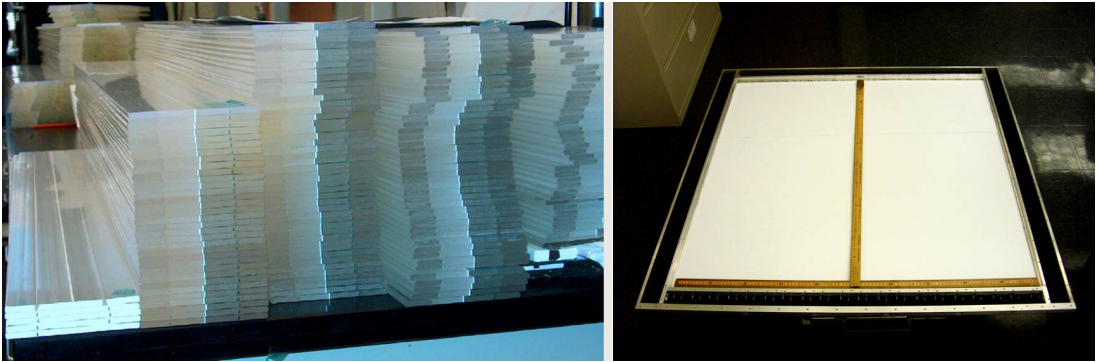}
\vspace{-6.0cm}
\caption{Left: TCMT scintillator strips after preparation and prior to assembly.  Right: TCMT Cassette during assembly. }
\label{fig:StripsandCassette}
\end{figure}

Scintillator strips form the active medium between the absorber plates and were assembled in 16 modules or cassettes. The cassettes were installed in the absorber stack with alternating vertical or horizontal strip orientations. Each cassette houses 20 extruded 5 mm thick scintillator strips that are 50 mm wide and 1000 mm long as shown in Fig. \ref{fig:StripsandCassette}.  The cassettes present a 1000~x~1000~mm$^2$ active sampling area to the beam, shown by the meter sticks on top of the cassette under construction in  Fig. \ref{fig:StripsandCassette}.  The TCMT strip complement totals 320.  

\begin{figure}[tbh]
\centering
\includegraphics[width=75mm]{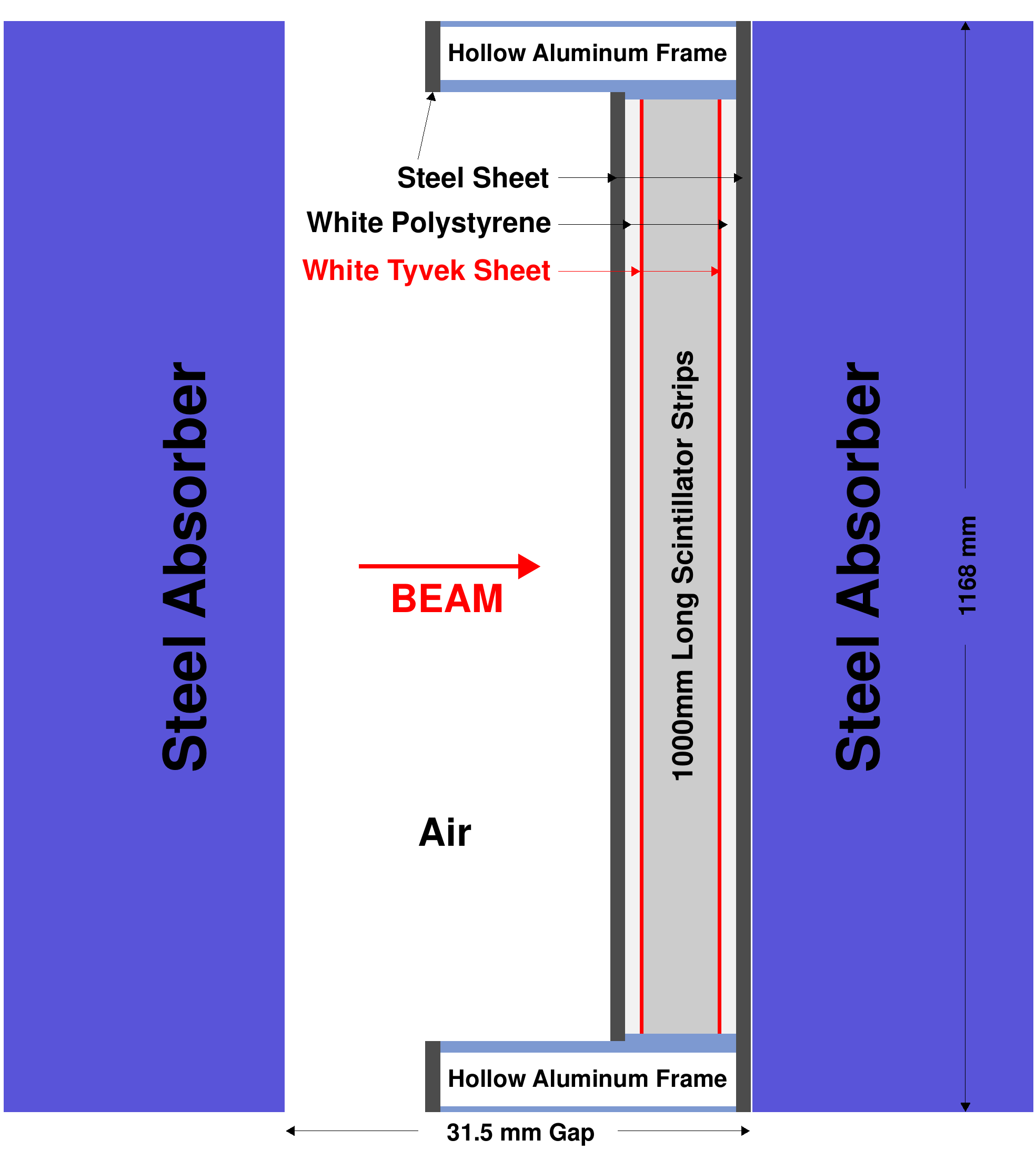}
\caption{Schematic of cassette cross section. SiPMs and LEDs are housed within the aluminum frame.  Horizontal and vertical scales are different, to clearly
  show cassette details.}
\label{fig:CassetteStructure}
\end{figure}

As shown in Fig. \ref{fig:CassetteStructure}, along the beam axis perpendicular to the face of the cassette, each cassette consists of a sheet of 1 mm steel, a single 1 mm sheet of white polystyrene plastic, a sheet of 0.15 mm Tyvek \cite{Tyvek}, the scintillator strips, another sheet of Tyvek, another sheet of white plastic, and the final rear steel sheet.  The layered structures are captured with a hollow rectangular aluminum frame extending  beyond all four edges of the square meter of active area. On one side of a cassette the hollow  frame houses SiPMs and on the opposite side the frame houses LEDs.   The steel sheets serve to mechanically protect the scintillator as well as provide a light-tight enclosure for the scintillator, the fibers, and the SiPMs.  As shown in the figure, the upstream steel sheet covers the surface area of the scintillator.  The downstream steel sheet covers the surface of the scintillator and the aluminum frame.  The aluminum frame is 20 mm thick with a 1 mm steel cover sheet or cap on the upstream side, ensuring the interior of the frame is light-tight.   The upstream steel cap, the aluminum frame, and the downstream steel sheet set the maximum width of the cassette to 22 mm.  The outer dimensions of the aluminum frame are 1168 x 1168 mm$^2$ to match that of the absorber.  As shown in the figure, the 22 mm cassettes are mounted adjacent to the downstream absorber and fit easily in the 31.5 mm gaps between the steel absorbers, the additional space provides tolerance for the absorber (which was unmachined) and clearance for assembly.   The inner square meter or active area of each cassette has a nuclear interaction length of 0.02 $\lambda_n$.  Thus, the total thickness of the sampling calorimeter from the beginning of the first cassette to the end of the sixteenth cassette (again not including the final steel layer) has a length of 1370 mm and a nuclear interaction length of 5.5 $\lambda_n$. 

The strips were developed and manufactured at the Fermilab/Northern Illinois Center for Accelerator and Detector Development (NICADD) extruder facility.  The scintillator was composed of polystyrene doped with 1.0\% PPO and 0.03\% POPOP.  The extruded scintillator had a width of 100 mm, a length of 1100 mm, and two 2 mm diameter co-extruded holes running the length of the scintillator.  The holes were 25 mm from each side. A 1000 mm long groove fully penetrating the scintillator material was cut down the center of each extrusion until 50 mm of material remained at each end of the scintillator. The groove was filled with a white epoxy made of DER 332 resin and Jeffamine \cite{Epoxy}.  This separation groove created two 50 mm wide strips. The ends of each extrusion were machined off, leaving a 1000 mm long element with a pair of two distinct strips separated by the white epoxy.  Sample strip pairs can be seen in Fig. \ref{fig:StripsandCassette}.

All scintillator strips were tested for uniformity of response.  Two strips at a time were placed in a light-tight box and both instrumented with a wavelength shifting fiber that was read out with the same photomultiplier tube.  The scintillator was excited with a Strontium-90 source and the photomultiplier output current recorded. The strips were subject to three criteria to ensure uniformity:  First, for each strip the average of measurements at 100, 200, 500, 800, and 900 mm from the edge nearest the photomultiplier were required to exceed a minimum threshold. Second, for each strip the ratio of the response at the 100 and 200 mm points, at the 200 and 800 mm points, and at the 100 and 900 mm points were required to be within two standard deviations of the mean ratio for the batch of strips. Third, the ratio of the average response of the two strips was required to be within two standard deviations of the mean ratio. With the above criteria, 227 pairs or 454 strips out of 468 strips qualified for assembly, corresponding to a 97\% yield.

Additional steps were taken to increase the light yield and improve uniformity of the strips.  The outer edges of the strips were painted with a reflective white paint, EJ-510 from Eljen Technology \cite{EJ-510}. The strips were not individually wrapped.   As mentioned earlier, the wider surfaces of each segment are adjacent to Tyvek sheets.   In this configuration measurements using the radioactive source revealed approximately 20\% attenuation between the ends of a strip.  Use of VM-2000 \cite{VM-2000}, a more reflective material than Tyvek, to cover a portion of the strips farthest from the photodetector reduced the attenuation to approximately 10\%.  The VM-2000 was actually glued to the Tyvek sheets. For each of five cassettes a 300 x 1000 mm$^2$  sheet of VM-2000 was positioned and glued onto one Tyvek sheet to cover  the strips in the region farthest from the sensor.  A second 200 x 1000 mm$^2$ sheet of VM-2000 was positioned on the other Tyvek sheet to cover the other side of the strips in the region farthest from the sensor. The dimensions of the VM-2000 sheets were selected after a series of systematic studies which measured the uniformity of response as a function of sheet area and position.  These five cassettes were installed as layers one, two, three, five, and seven of the TCMT, where layer one is nearest the beam source.

Light in each strip was collected with 1.2 mm diameter, 1035 mm long, Kuraray Y11 wavelength shifting (WLS) fiber with a polished and aluminum-mirrored end protected with UV cured epoxy. The uniformity of the fiber response was better than 1.0\% \cite{Fibers}.  The fiber directed the light to a SiPM (supplied by the Moscow Engineering and Physics Institute (MEPHI) \cite{SiPM} in collaboration with Pulsar Enterprise) for detection. 

The SiPMs were part of the same production runs utilized for construction of the CALICE scintillator/SiPM hadronic calorimeter.  The operation and performance of the SiPMs are fully discussed in the reference describing the hadronic calorimeter\cite{AHCAL}. As mentioned above, photons collected by the fiber are shifted from the blue wavelength of the scintillator to the more sensitive green region of the SiPM.  The SiPMs are all mounted on one side of the cassette with custom made holders for the WLS fiber and the SiPM.  The fiber-SiPM air gap was held within 0.1 mm and transverse displacement was within 0.1 mm. The SiPMs are connected with Radiall \cite{Radiall} multi-coax connectors to the front-end electronics via 50 $\Omega$ coax cables that carried the bias voltage on the shield. 

For calibration and monitoring, each strip was illuminated with a light emitting diode (LED) driven by a LED driver circuit installed in a cassette panel opposite the SiPM sensors. The LEDs emit light at wavelengths (blue and ultraviolet) consistent with scintillation light in the plastic scintillator.  Each TCMT cassette has a 20 channel LED Driver Board (LEDDB).  Each group of four cassettes is connected to a custom LED Fanout Board (LEDFB) each with an independent power supply; a total of four LEDFBs were needed for the entire TCMT. 

The LEDFBs and LEDDBs are controlled by the DAQ system.  During calibration, the DAQ transmits an analog signal to each LEDFB where the timing is adjusted for the SiPM readout and then relayed to the LEDDBs.  The signal activates all LEDs for a 20 ns illumination interval. During cassette assembly, individual LEDDB channels were adjusted so that the response of all 20 LEDs in a single cassette to a single external signal were similar.  Since the DAQ could only output a single voltage level per eight cassettes, a scan of the level was performed to find the level generating the maximum number of useful photo-electron spectra. The lack of an independent level for each individual channel resulted in channels without calibration spectra; either the signal was too high, saturating the spectra, or too low, without clear photo-electron peaks.

\section{Installation, Beam Line, and Data Collection}
\label{sec:Section3}

\begin{figure}[tbh]
\centering
\includegraphics[width=150mm]{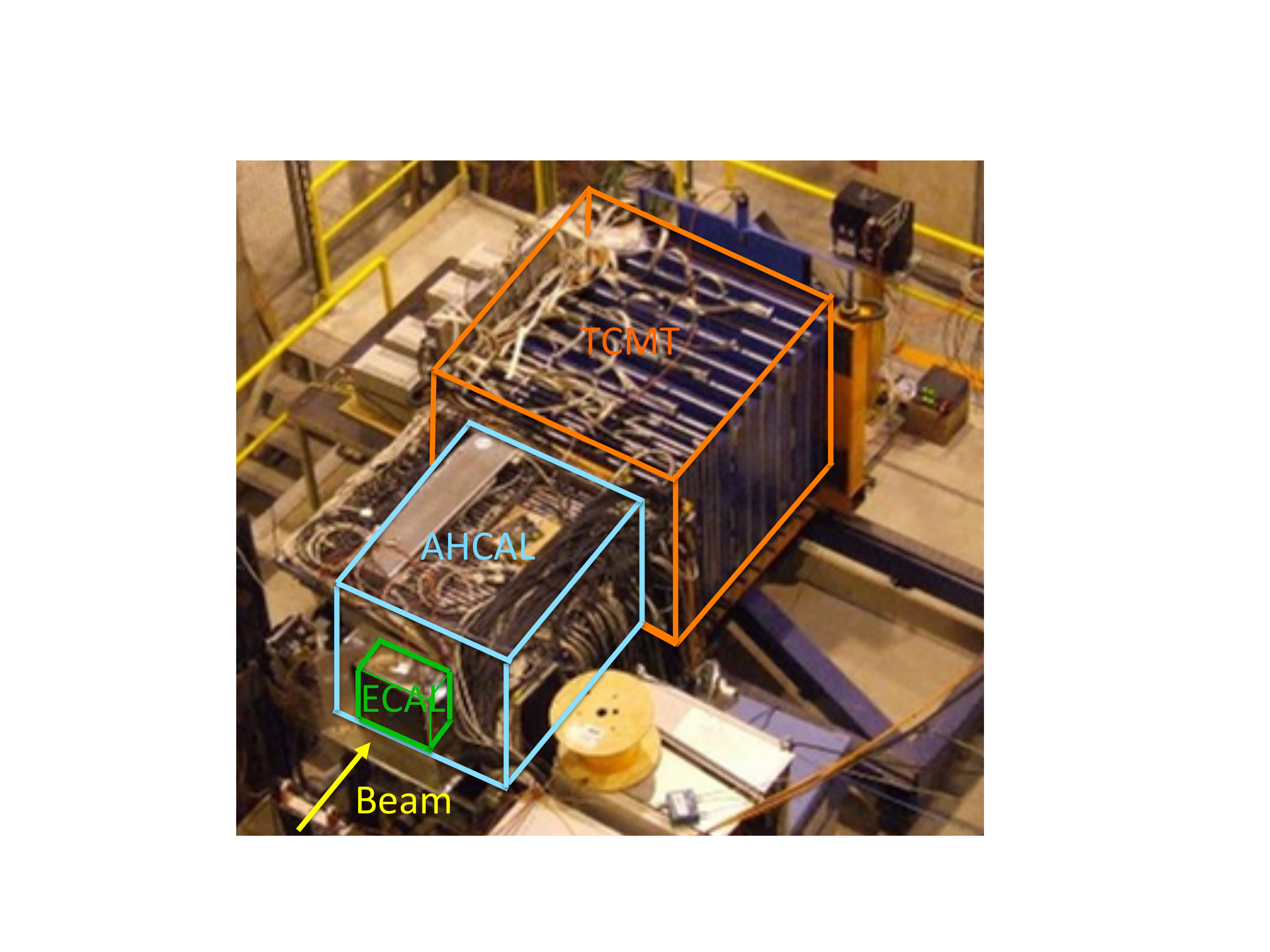}
\caption{The CALICE detector at the 2006 CERN testbeam. The TCMT is traced by the labeled orange outline.}
\label{fig:TestbeamTCMT}
\end{figure}

In August 2006 the TCMT was installed with the CALICE silicon tungsten electromagnetic calorimeter (ECAL) \cite{ECAL} and the analog hadronic calorimeter (AHCAL) \cite{AHCAL} in the H6 beam line at the CERN SPS.  Figure \ref{fig:TestbeamTCMT} shows the CALICE calorimeters fully assembled on a translating platform.  The TCMT is the large module near the rear of the photograph; the aluminum/steel cassette surfaces are visible between the blue absorber plates.  

The silicon-tungsten ECAL prototype contains 30 layers assembled into three modules. The tungsten absorbers have different thickness for each module (1.4, 2.8, and 4.2 mm with the thinner absorbers nearest the beam origin).  The silicon planes are divided into 3x3 wafers, each wafer having 6x6 cells of 10 x 10 mm$^2$.  For the October 2006 run, each plane had six out of nine wafers in place for a total of 6480 channels.  Details are available in reference \cite{ECAL}.

As designed, the AHCAL includes 38 steel and scintillator layers, each layer with a volume of
900.0 x 900.0 x 32.0 mm$^3$.  Each layer is a combination of a 17.4 mm thick steel absorber plate and a 11.7 mm thick readout module or cassette. In addition each layer has a 2.5 mm air gap to allow for tolerances.   The AHCAL cassette is composed of two 2 mm steel covers, 5 mm thick scintillator segmented into 216 tiles read out with WLS fiber and SiPMs, a 1 mm printed circuit board, 1.5 mm thick layer of cables and fibers, and 0.2 mm of foil.   The full AHCAL has 7608 electronic channels.  See the references for details \cite{AHCAL}.  The AHCAL cassettes differ significantly from the TCMT cassettes primarily because the SiPMs and associated cables are internal to the cassette.

The data discussed in this paper were collected in October 2006, when the AHCAL was composed of 30 layers of steel with gaps one through 17 instrumented with a cassette and every \textit{other} gap between 18 and 29 (starting at 19) instrumented with a cassette.  The total of 23 cassettes corresponds to 4968 channels.  

\begin{figure}[th]
\begin{center}
\includegraphics[width=130mm]{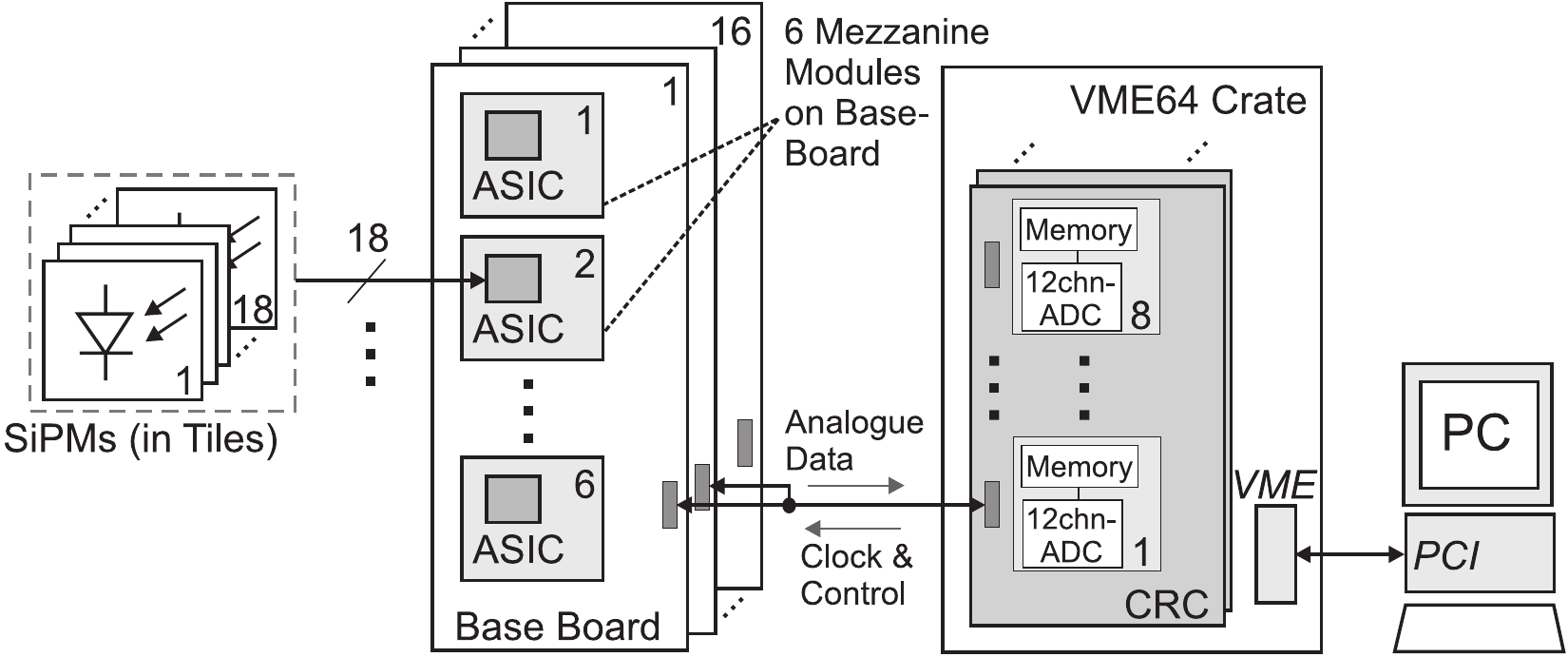}
\end{center}
\caption{Schematic view of the CALICE readout and data acquisition.}
\label{fig:DAQ}
\vskip 0.2cm
\end{figure}

The TCMT readout system was fully integrated with the CALICE calorimeter readout system described in detail in reference \cite{AHCAL}.  Briefly, as shown in Fig. \ref{fig:DAQ}, the CALICE data acquisition system (DAQ) system is based on a VME crate which houses CALICE Readout Cards (CRC) carrying memory and the ADC converters for the SiPM signals.  Each CRC supports 16 AHCAL base boards (HBAB).  Each HBAB has six mezzanine boards, called the AHCAL analog board (HAB), with an ASIC which can accommodate 18 SiPM channels.  The HAB provides an output driver for the SiPM signal, control and configuration electronics set remotely by the DAQ, and SiPM bias voltage. Not shown are adapter boards between the TCMT cassettes and the HBABs which translate the 20 channels per cassette to the 18 channels per mezzanine boards.  There is one adapter board per HBAB.  A total of one CRC, four HBABs, 24 HABs, and four adapters comprising 432 channels were reserved for the TCMT although only 320 channels were needed.  The DAQ also controls LED operation through the LEDFB. 

The DAQ system was controlled by C++ software, developed by the CALICE collaboration, running on a PC connected to the VME crate.  The stored information contains the CALICE detector data (including hardware configuration) as well as information from beam elements.  The data was written in a custom binary format and converted offline to the LCIO format (the standard data format for ILC related studies) \cite{LCIO}.  The full system could operate at 120 Hz.  Data from each run was recorded in a series of files containing up to 200k events.  Every five minutes during each run 500 event pedestal calibration data sets were taken followed by a 500 event LED calibration run.  Further details are given in reference \cite{AHCAL}.

\begin{figure}[tbh]
\centering
\includegraphics[width=90mm]{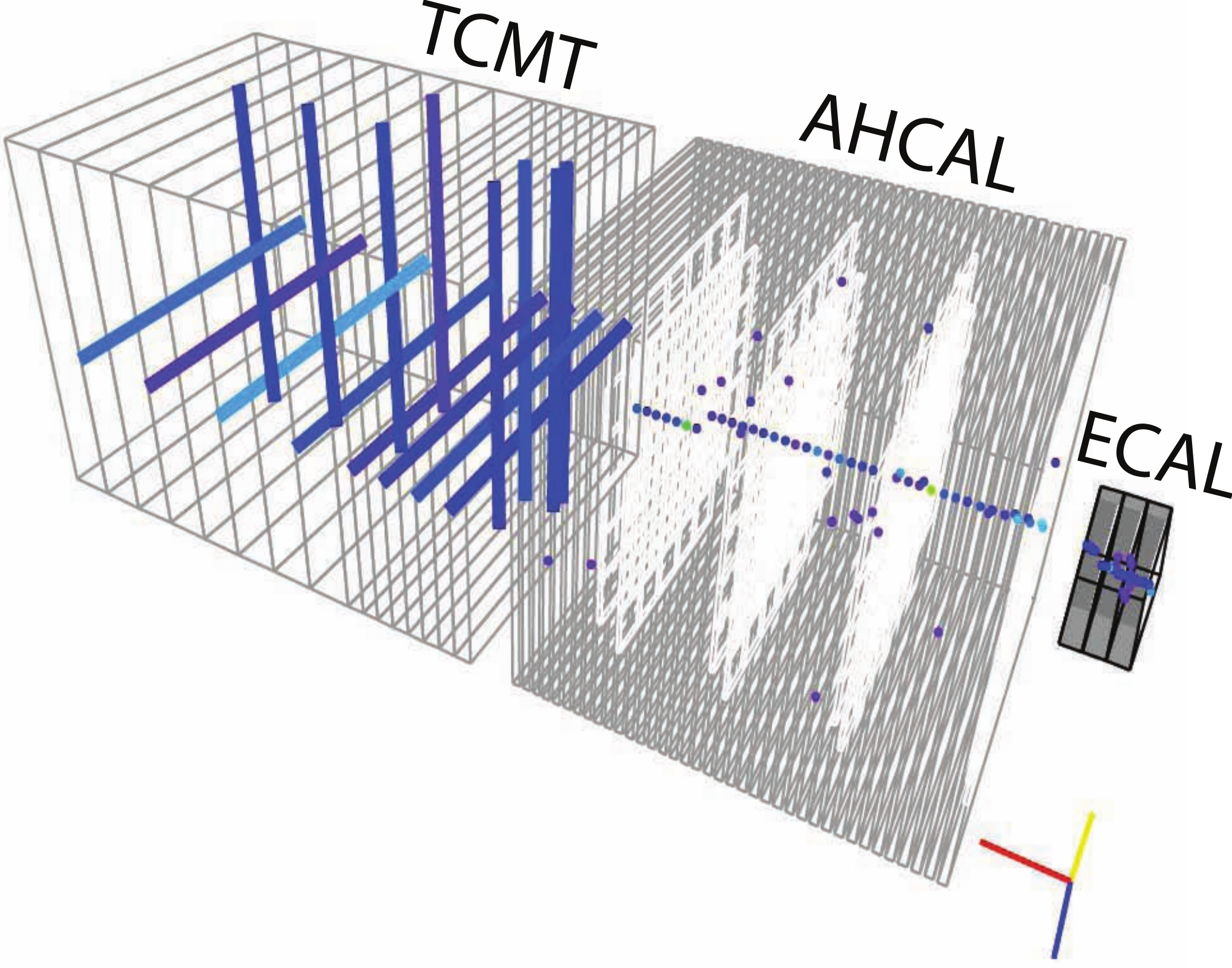}
\caption{Muon event recorded at the CERN testbeam.  Hits above threshold are highlighted. }
\label{fig:muonevent}
\end{figure}

Data was collected for electrons, pions, and muons. Approximately 17 million pion events and 10 million muon events were collected.  Figure \ref{fig:muonevent} shows a muon passing through the CALICE calorimeter; the TCMT strip hits are visible.  Table \ref{tab:TableI} lists the pion runs with their momentum, p, and the associated numbers of events taken in October 2006 and analyzed in this paper.  A full description of the 2006 data run can be found in reference \cite{AHCAL}.  Although each beam run had a dominant particle species, contamination was present.  For example, a 20 GeV negative pion run was contaminated by as much as 20\% non-pions. 

\begin{table}[tbp]
\centering
\caption{Momentum and charge of pion events collected in October, 2006.}
\vspace{0.1cm}
\begin{tabular}{|c|c|c|c|c|c|c|c|c|c|c|c|}
\hline
p(GeV) & 6 & 8 & 10 & 12 & 15 & 17 & 20 & 30 & 40 & 50 & 80 \\
\hline
$\pi^{+}$($10^{6}$events) & 0.45  &          &  0.70 &          &  0.70 &          & 0.80  & 0.80 & 0.80 & 1.30 & 1.30 \\
$\pi^{-}$($10^{6}$events)  & 1.40 & 1.30 & 1.80 & 1.20 & 1.60 & 1.30 & 1.50 &       &       &          & \\
\hline
\end{tabular}
\label{tab:TableI}
\end{table}

\begin{figure}[tbh]
\centering
\includegraphics[width=130mm]{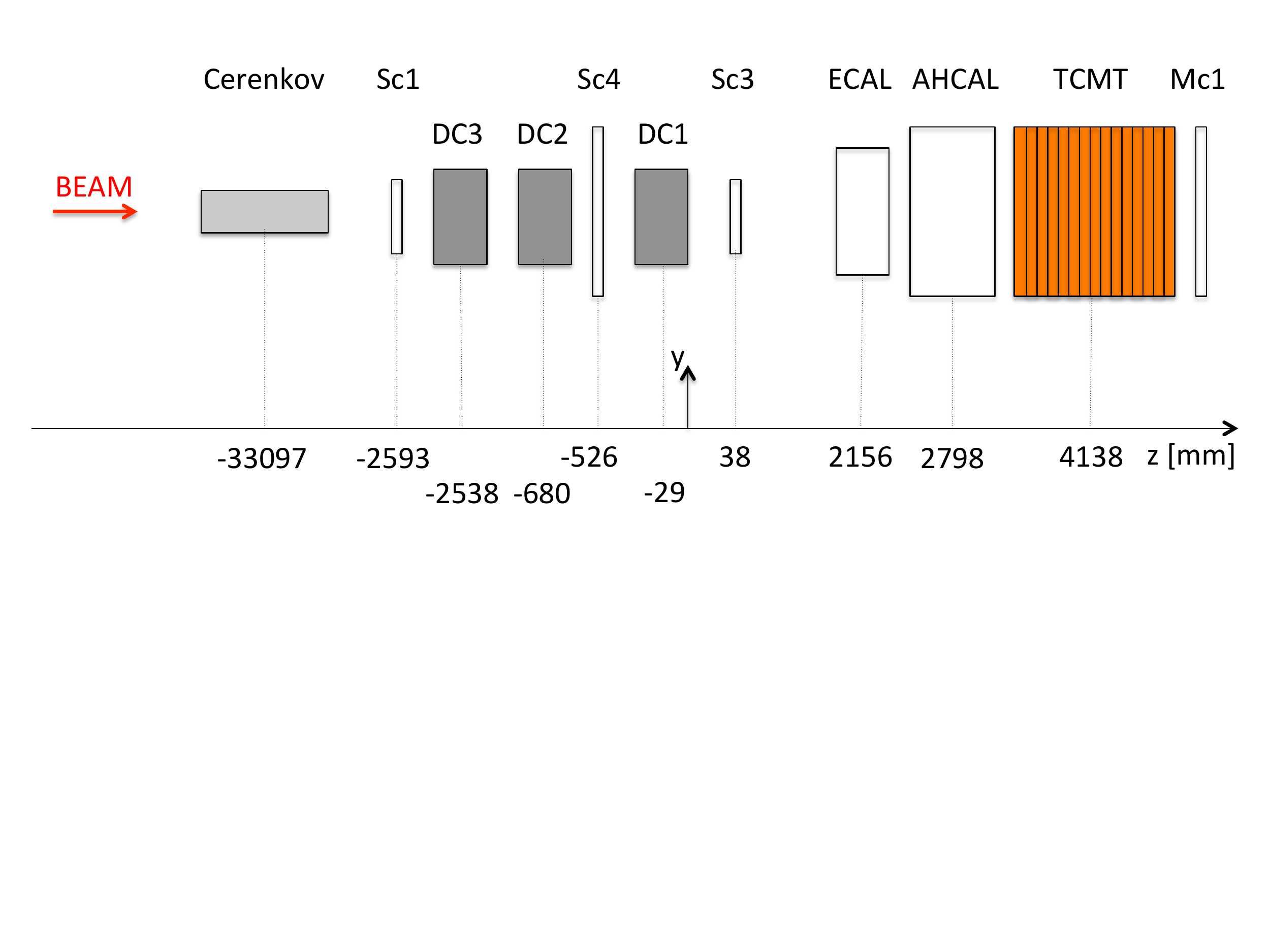}
\vspace{-4.0cm}
\caption{October 2006 CALICE beam line components.  Not to scale.}
\label{fig:Beamline}
\end{figure}

A schematic of the beamline configuration and components during October 2006 is shown in Fig. \ref{fig:Beamline}. The readout was triggered by a coincidence of the two 100 x 100 mm$^2$ scintillator counters labeled Sc1 and Sc3. A larger scintillator Sc4, of dimensions 200 x 200 x 10 mm$^3$ was used offline to reject double particles or showers initiated in the air or material before the detector.  Beam particle tracking was provided by three drift wire chambers DC1, DC2, and DC3.  Muon identification was provided by Mc1, a 1000 x 1000 mm$^2$ scintillator counter, located behind the calorimeter elements.  A Cerenkov counter identified electrons and pions.  The response of the beam components was recorded for offline use.  Scintillator counter amplitudes were also used to indicate the presence of multi-particle events offline. 

The muon counter Mc1 was constructed at NICADD and similar to the TCMT cassettes.  Mc1 was used as an on-line trigger for calibration muons and off-line as a muon veto.  The counter was made of twenty 1000 x 50 mm$^2$ extruded strips of 50 mm thickness with a co-extruded hole in which a Kuraray Y11, 1.2 mm diameter, WLS fiber was embedded.  To ensure uniformity of response all 20 WLS fibers were of the same length.  To reduce the impact of cladding light from the fiber, each fiber extended 0.7 m beyond the end of the strip.  The fibers were connected to a single photomultiplier tube through a long square crystal. The scintillator end of each WLS fiber was polished, then mirrored with aluminum, and protected with UV cured epoxy. The photomultiplier end of each fiber was glued into a square plastic ferrule made from Delrin and finished with the fly diamond cutting technique.  The square ferrules simplified assembly and made the fiber connection to the crystal robust.

\section{Stability, Calibration, and Light Yield}
\label{sec:Section4}

The initial performance of the TCMT was explored with random trigger (pedestal), LED calibration, and muon runs.  The pedestal runs provide information on the stability of the channel.  Muon runs are used to calibrate each channel.   In addition, the strip light-yield was measured with the LED system and the muon runs.  Of the 320 channels, 312 or 97.5\% operated during the test beam run.  No breakage was encountered after installation. 

\begin{figure}[tbh]
\centering
\includegraphics[width=110mm]{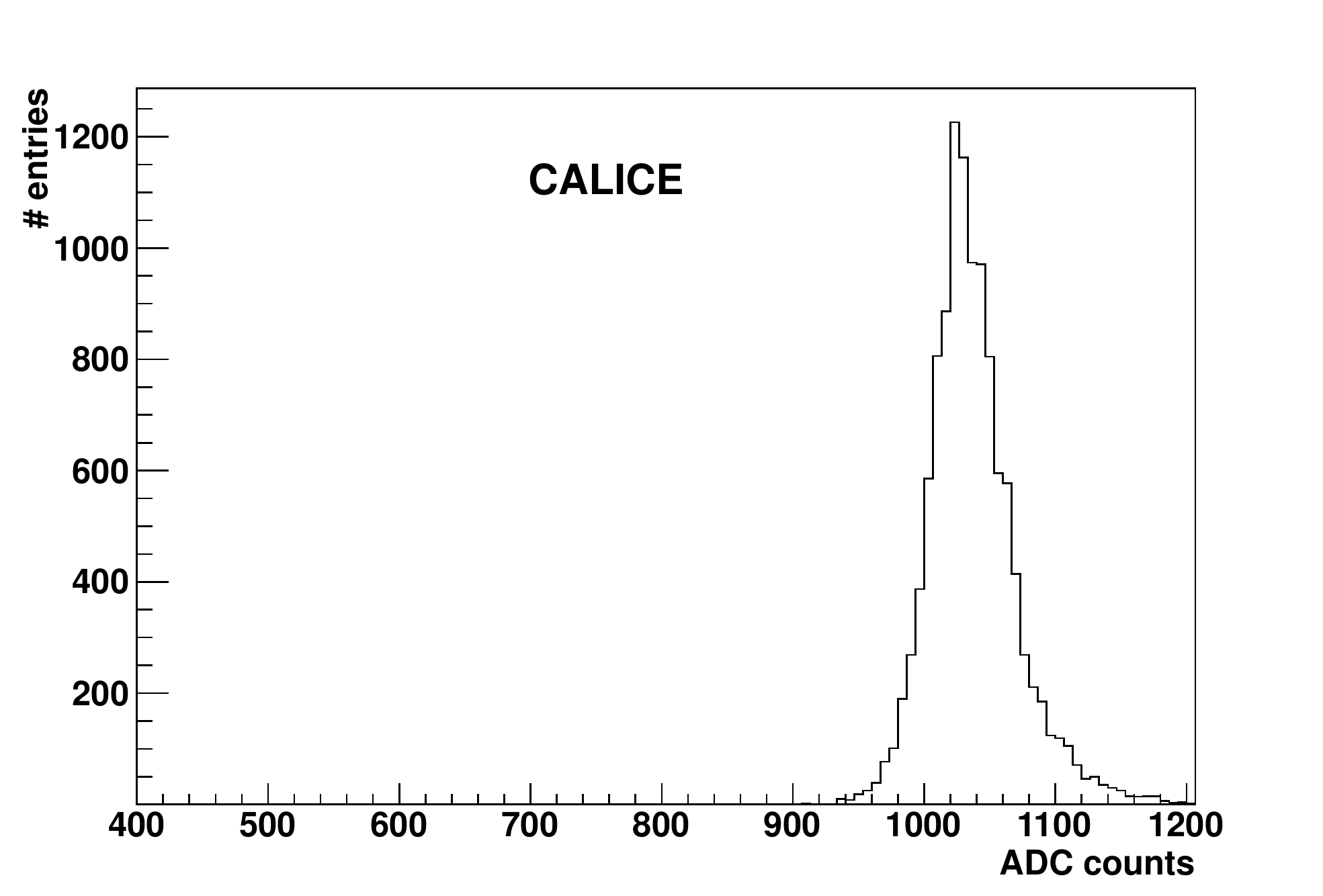}
\caption{Single strip pedestal or noise distribution collected over several pedestal data samples.}
\label{fig:Pedestal}
\end{figure}

\begin{figure}[tbh]
\centering
\includegraphics[angle=90, width=110mm]{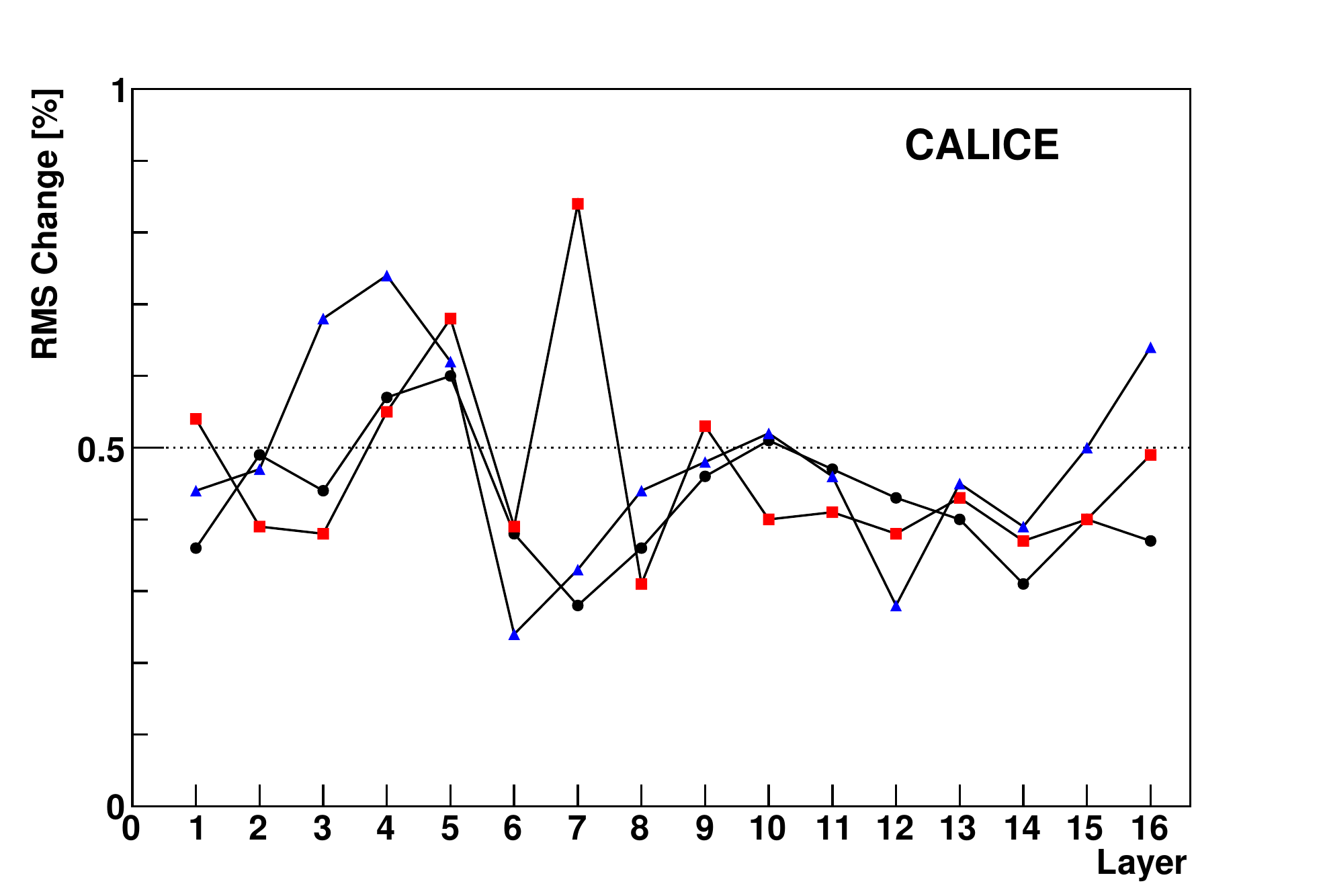}
\caption{Percent variation of the pedestal root-mean-square  for all strips in each cassette (layer) for three runs relative to an initial run taken over a one day period.}
\label{fig:Stability}
\end{figure}

Pedestal stability was investigated with the regularly accumulated 500 event pedestal samples.  An example pedestal distribution for a single strip is shown in Fig. \ref{fig:Pedestal}; the statistical mean was taken as the pedestal for each strip.   The time dependence of the pedestal mean and width provides information on the stability of the ground level and of the noise of the channels, respectively.  A typical pedestal was stable to about 9 ADC counts or 700 $\mu$V over a 24 hour period.  As shown in Fig. \ref{fig:Stability}, the change of pedestal root-mean-square (RMS) for the 20 channels in each of the 16 cassettes between four runs taken over a 24 hour period was stable to less than one percent.

\begin{figure}[tbh]
\centering
\includegraphics[width=110mm]{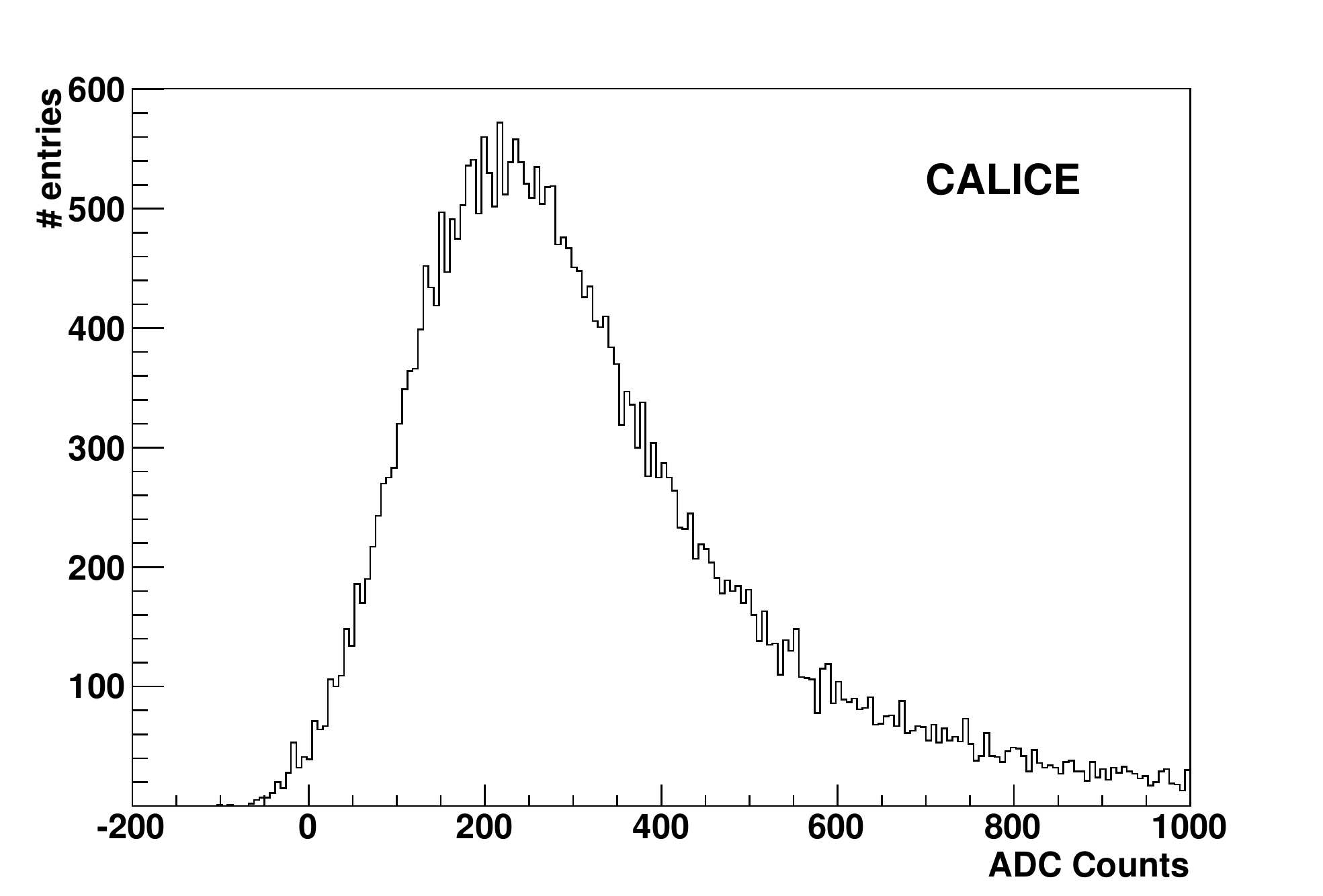}
\caption{Single-strip response to muons. Events were selected by requiring hits above pedestal in corresponding strips in neighboring cassettes.}
\label{fig:MuonSpectrum}
\end{figure}

Figure \ref{fig:MuonSpectrum} shows the pedestal-subtracted response of a sample strip to a muon beam incident on the strip.  In addition to the on-line muon trigger requirements, these muon events were selected offline by requiring a signal 2$\sigma$ above pedestal in five of six nearest neighbor cassettes.  Wherever the layer position permitted, three strips in front and three strips behind were required.  The response in ADC counts was fit with a Gaussian distribution to the left of the peak value and over a limited range to the right.  The  Gaussian distribution is used to calibrate each strip
for later analysis such that the mean (or the most likely ADC value) after pedestal subtraction is defined to be one in units of minimum ionizing particles (MIP).   The mean in Fig. \ref{fig:MuonSpectrum} is around 225 ADC counts, thus the 9 ADC count pedestal drift corresponds to approximately 0.05 MIP.  Although a combined Gaussian distribution and Landau distribution would better accommodate the high tail, the procedure used to calculate weighting functions for each TCMT layer (described later) minimizes any impact on particle energy measurements from the limited Gaussian fit.  In addition, any fluctuations induced by the specific fitting function are small relative to the stochastic variation in pion showers.

\begin{figure}[tbh]
\centering
\includegraphics[angle=90, width=110mm]{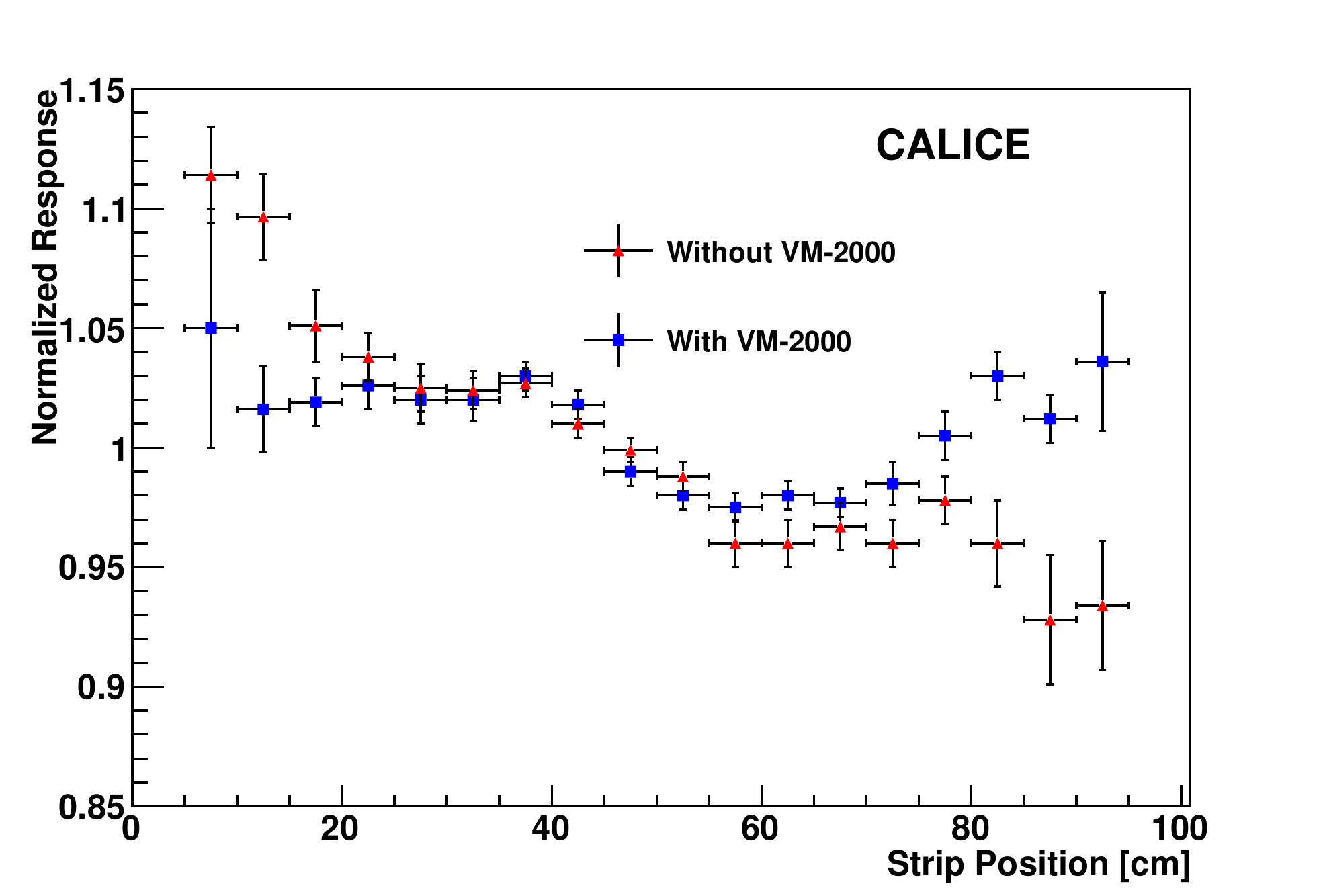}
\caption{Normalized weighted response of three strips with VM-2000 treatment (blue square symbols) and three strips without  treatment (red triangular symbols) to muons as a function of position along the strip. Response was normalized to unity near the strip center.  The horizontal uncertainty bars correspond to the strip width.}
\label{fig:MuonAttenuation}
\end{figure}

Figure \ref{fig:MuonAttenuation} shows the weighted response to muons of three strips with additional VM-2000 treatment and three strips without additional treatment as a function of muon beam position.  Both sets of strips were near the center of the TCMT (strips 9, 10, and 11).   The response was normalized to unity at strip center.  Strips of opposite orientation in adjacent cassettes were used to locate the position of muon entry.  A linear fit to the untreated strip response yields an attenuation of  (17.6$\pm$1.3)\% per meter, consistent with measurements (discussed earlier) using a radioactive source.   As expected, the VM-2000 treated strips show less variation in response.   

After calibration with muons, crosstalk between strips was measured at 4.5\%  by comparing the response in strips adjacent to a strip with a muon signal between 1.5 and 4.0 MIPs.  Since the strips are not individually wrapped but are covered by sheets of reflector, light can leak from one strip to another. The magnitude of the crosstalk was calculated with two methods.  First, the data from all adjacent strips were combined in a single plot; second, each strip was examined independently.  Consistent results were obtained with the two methods.  The level of cross-talk was also consistent with measurements with a radioactive source and was qualitatively described with a  GEANT-based simulation in which cross-talk was modeled by redistributing energy deposited in a strip to neighboring strips.

\begin{figure}[tbh]
\centering
\includegraphics[angle=90, width=110mm]{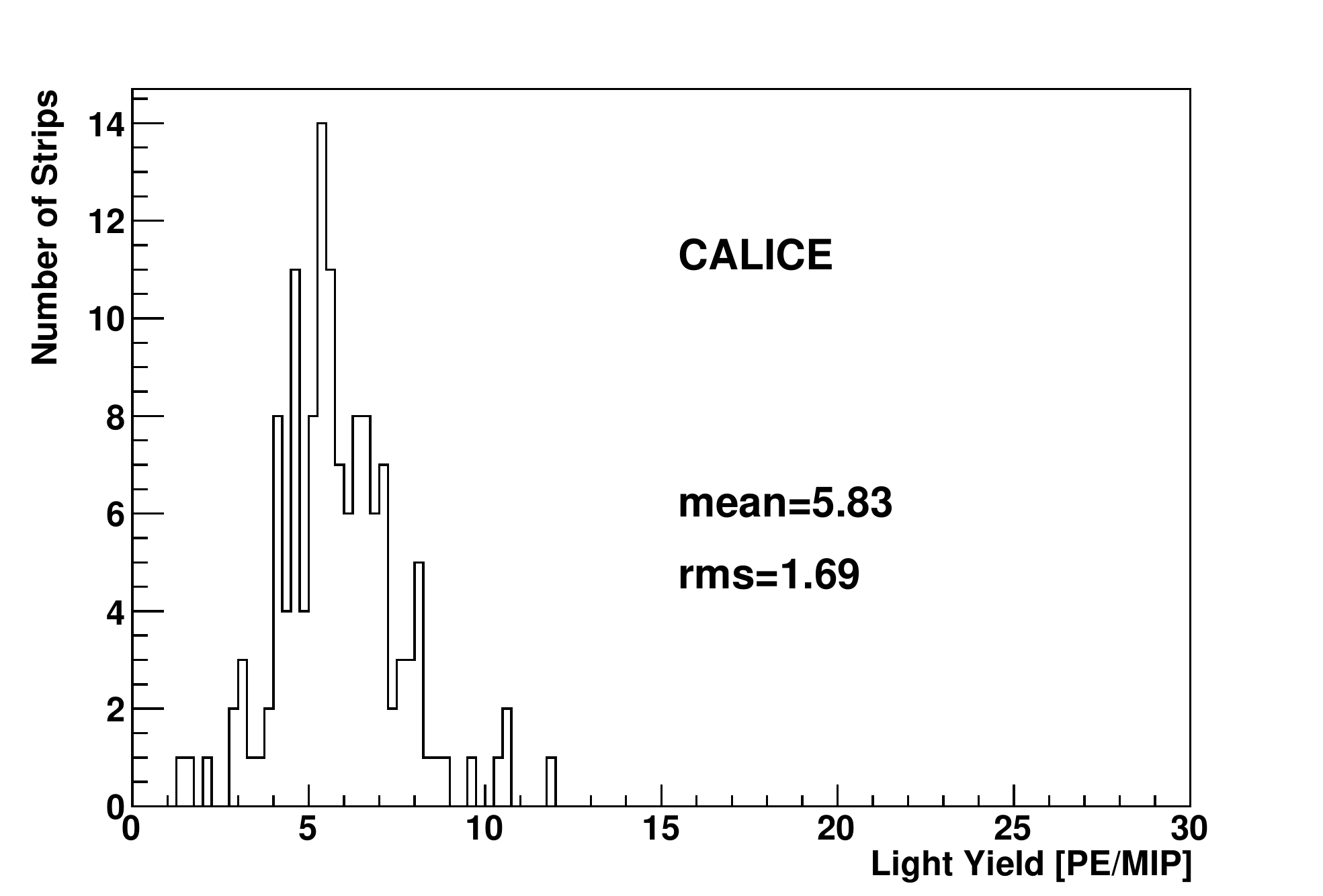}
\caption{Distribution of strip light yield in photoelectrons.}
\label{fig:LightYield}
\end{figure}

Determination of the light yield in the strips required SiPM gain calibration with the LED system. In LED calibration runs, each channel was illuminated with a LED and the photo-electron spectrum recorded.  (See the references for example spectra \cite{AHCAL}.) The separation between the first and second photo-electron peaks in ADC counts provides a measure of the SiPM gain in units of PE/ADC.  This calibration gain multiplied by the MIP calibration in units of ADC/MIP provides a measure of the light yield in PE/MIP.  The LED calibration data and the muon beam data were collected with different amplifier gains and shaping times to increase dynamic range, minimize noise during data taking, and to ensure well separated PE peaks in the calibration data.   For each channel a correction factor, typically about ten, was determined by comparing the mean response of LED runs taken in the calibration (high gain) and beam (low gain) modes. See reference \cite{AHCAL} for details.  The light yield was calculated for only 135 channels due to limitations of the LED calibration system (as discussed in Section 2). 
As shown in Fig. \ref{fig:LightYield} the extruded-strip light yield in terms of photo-electrons per minimum ionizing particle (PE/MIP) has a mean of 5.83 PE/MIP and an RMS of 1.69 PE/MIP.  

\begin{figure}[tbh]
\centering
\includegraphics[width=110mm]{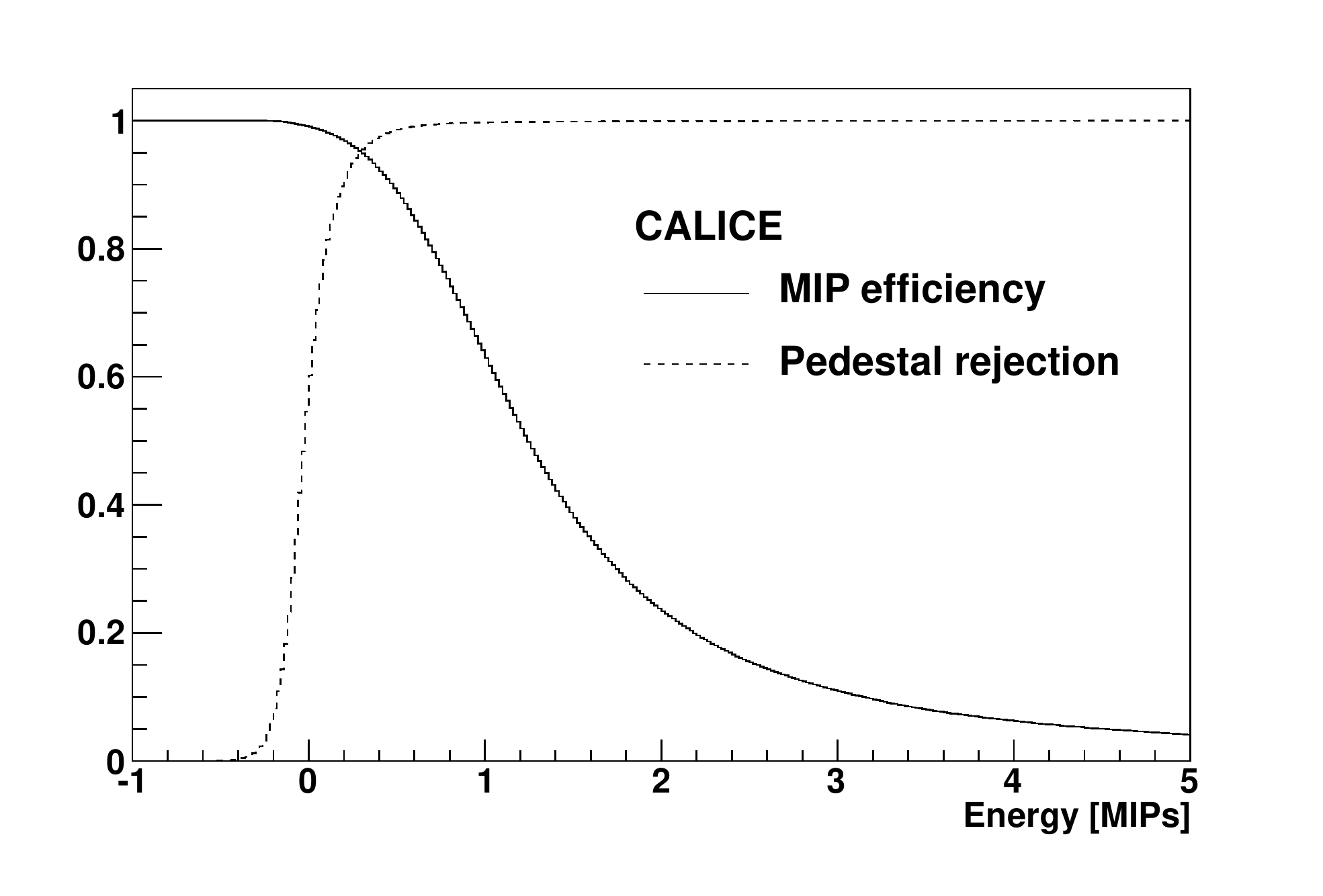}
\caption{Pedestal rejection (dashed curve, increasing from zero to unity with energy) and MIP efficiency (solid curve, decreasing from unity to zero with energy) as a function of threshold energy in MIPs.}
\label{fig:Threshold}
\end{figure}

To suppress noise in the analysis, the response for each strip after pedestal subtraction must exceed a minimum threshold.  
Figure \ref{fig:Threshold} shows pedestal rejection (the fraction of pedestal events exceeding a threshold) and muon detection efficiency (the fraction of muon events exceeding the same threshold) as a function of threshold for an example TCMT channel.  
In the figure the "cross-over" point where the two curves are equal corresponds to $\sim$95\% pedestal rejection and  $\sim$95\% muon efficiency.  Analysis of all active strips shows that at a threshold of 0.5 MIP the pedestal rejection is (97$\pm$3)\% and the MIP efficiency is (85$\pm$3)\%. The efficiency may be slightly under-estimated near zero MIPs due to residual pedestal contamination in the muon calibration runs, however visual inspection indicates this is very small.  For further analysis a threshold of 0.5 MIP was required for each strip.  
 
\section{Hadronic Response}
\label{sec:Section5}

The pion data collected in 2006 was used to characterize the response of the TCMT to hadronic showers.  After event selection, correction for saturation of the SiPMs, and optimization of sampling weights the relative contributions of the TCMT to the overall energy resolution of the CALICE calorimeter in the October 2006 configuration was measured.  The TCMT also provides an opportunity to measure the leakage of energy from the ECAL and AHCAL.  Finally, the TCMT component of the full calorimeter system can be used to emulate the placement of a magnet coil in a detector, enabling us to study the effect of coil material on the energy resolution of a realistic ILC calorimeter.  This will be covered in more detail in Section 6.

\begin{figure}[tbh]
\centering
\includegraphics[angle=90, width=110mm]{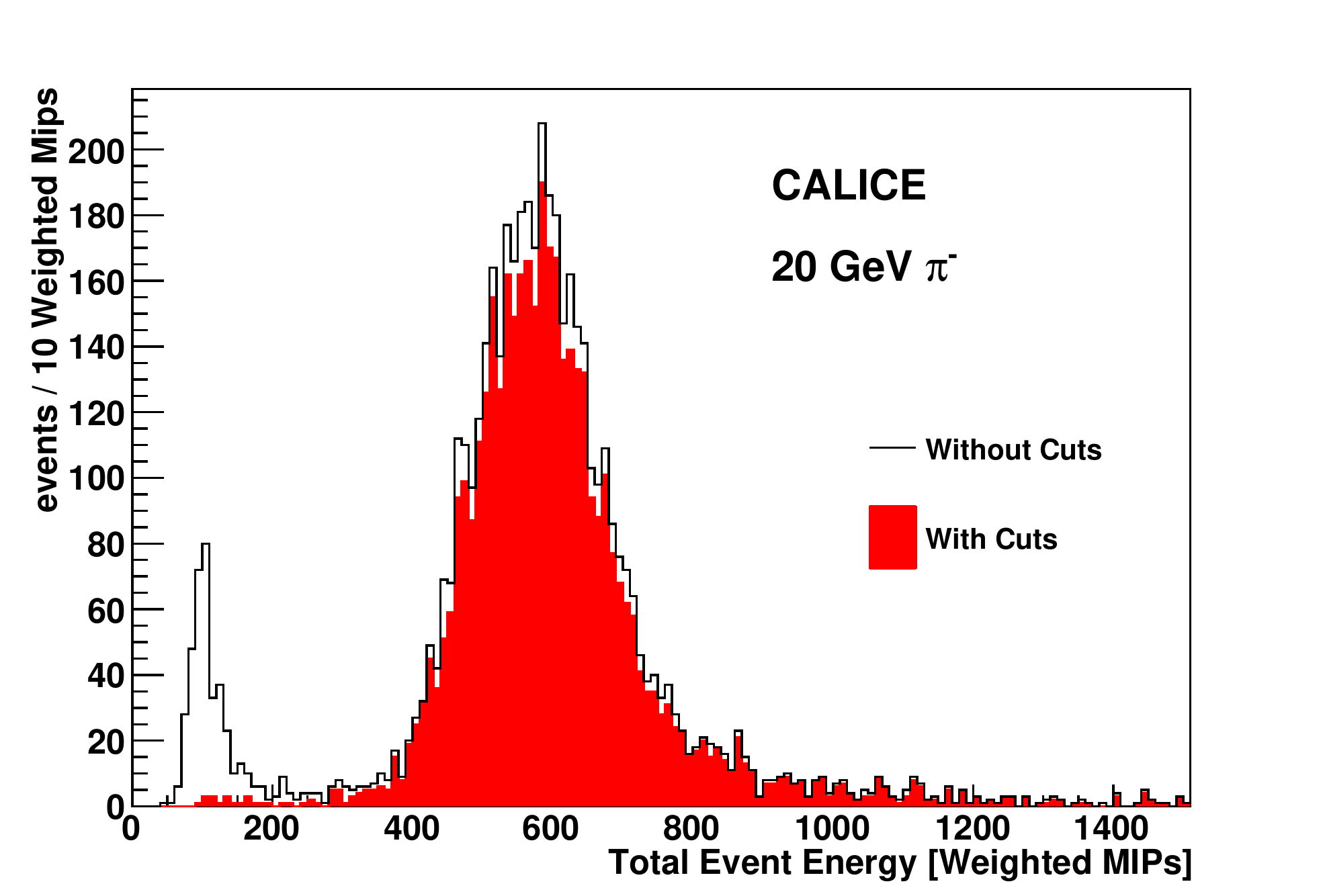}
\caption{The sum of energy in the ECAL, partial AHCAL, and TCMT for a pion sample without (open histogram) and with selection criteria (filled histogram, mean=617 MIPs, RMS=186 MIPs) to remove contamination.  Muons are evident in the lower energy peak. }
\label{fig:withandwithoutcuts}
\end{figure}

The pion runs included electrons, muons, and pions.  A number of on-line and off-line requirements ensure that the electron and muon contamination was reduced to a negligible level.  The open histogram in Fig. \ref{fig:withandwithoutcuts} shows the total ECAL, AHCAL and TCMT energy in units of MIPs for a 20 GeV negative pion run.  Muon contamination is evident in the peak near 100 MIPs; there are also multi-pion events in the high tail.
(The weighting of the contributions from the ECAL, AHCAL,  and TCMT was based on sampling fractions calculated between the absorber and active medium.)
Electrons are vetoed with the Cerenkov counter, multi-particle events are vetoed with a threshold cut in the scintillator counter (SC4 in Fig. \ref{fig:Beamline}) mounted before the calorimeters, and muons are vetoed with the scintillator counter behind the TCMT (Mc1 in Fig. \ref{fig:Beamline}) as well as vetoing on energy depositions in the ECAL, AHCAL, and TCMT consistent with a MIP.   The filled histogram in Fig. \ref{fig:withandwithoutcuts} shows the purified pion spectrum.  Typically, after cuts about 80\% of the events are retained.

As discussed elsewhere (~\cite{AHCAL}, \cite{Saturation}), because each SiPM has a limited number
of pixels (1156 in total), the response must be corrected for saturation. The saturation correction
occurs in three steps:   First, using the response gain derived by calibration with the LED board, the observed signal in a channel is converted to a light yield, given as a number of photons. Second, a saturation correction is applied which is is based on test bench measurements (by the manufacturer) of the number of SiPM pixels fired as a function of LED light intensity. At low light levels the number of pixels fired is linear with respect to intensity. Extrapolation from the linear region provides a set of correction values for higher illumination levels.  Finally, the gain is re-applied to the corrected light yield. For those channels for which the gain is unavailable, the gain was set equal to the average gain of those channels for which the gain was available. This introduces negligible error as the overall saturation correction is small for the TCMT. 

The correction is important in the AHCAL \cite{AHCAL} as the particle showers have regions of intense energy deposition, but much less so for the TCMT as the showers mostly develop well before the TCMT.  In the TCMT, for 10 GeV and 80 GeV pions 1.2\% and 5.8\% of the strip hits exceed the SiPM linear range, respectively.  Correction for the saturation increases the overall energy at 10 GeV and 80 GeV by 1.1\% and 2.1\%, respectively \cite{KFThesis}.

Sampling weights for each instrumented layer of the ECAL, AHCAL, and TCMT can be derived to optimize the response of the complete calorimeter. The sampling weights can be calculated for an event sample using a least squares minimization of the quantity
 \begin{center}
$\displaystyle\left(\frac{1}{N}\right)\sum\limits_{j=1}^N (E_{0}-\sum\limits_{i=1}^{n}A_{i}L_{ij})^{2}$
\end{center} 
where $N$ is the number of events in the data sample, $E_{0}$ the beam energy, $n$ is the number of distinct sections of the detector with a constant sampling fraction, $A_{i}$ is the weight of the $i^{th}$ section and $L_{ij}$ the signal in the $i^{th}$ section for the $j^{th}$ event.   The $A_{i}$ provide an optimized conversion from the number of MIPs to energy in GeV.  The sections can be individual layers or groups of layers.   For a fixed set of events the summation can be rewritten in matrix form and the weights uniquely determined via matrix inversion \cite{KFThesis}.  

To evaluate the hadronic energy resolution of calorimeter configurations of increasing depth or configurations emulating a magnet coil, 30 different sets of weights were required depending on the number and configuration of the TCMT layers used.  There are three distinct sections of the ECAL, two distinct sections of the AHCAL, and two distinct sections of the TCMT, each characterized by different absorber thicknesses. Five to eight weights were used depending on the configuration under investigation. As an example, for a calorimeter which only includes the three sections of the ECAL, corresponding to the three absorber thicknesses, and the two sections of the AHCAL, five weights are required. The first section of the AHCAL includes the 18 fully instrumented steel/scintillator layers. The second section corresponds to the twelve absorber layers in which every other layer was instrumented with scintillator.  Configurations which extended the calorimeter with the TCMT, had five to seven weights: five when only the ECAL and AHCAL were used, six when the TCMT layers with thin absorber are added and seven when both thin and thick TCMT absorber sections were added. 

Table \ref{tab:TableII} shows the weights derived from a 20 GeV negative pion run for six of the 17 configurations in which successively more TCMT layers are added to the calorimeter system. \mbox{Configuration 0} has no TCMT layers and Configuration 16 has all 16 TCMT cassettes added. Columns ECAL1, ECAL2, and ECAL3 represent the weights for the three ECAL sections differing by absorber thickness. AHCAL1 represents the first 17 layers of the AHCAL in which an active layer was placed after each absorber. AHCAL2 represents the section of the AHCAL in which active layers were installed between every other absorber. The TCMT1 column represents the weight for the eight TCMT layers with a 19 mm thick absorber. The TCMT2 column represents the weight for the eight TCMT layers with a 102 mm absorber. 
No requirements were placed on the energy deposition in the ECAL, showers could have developed early in the calorimeter.  As can be seen, the ECAL and AHCAL1 section weights vary little between configurations.
The greatest variations between configurations are seen in the TCMT weights as the measured correlations between layers change as layers are added to the TCMT. 

\begin{table}[tbp]
\centering
\caption{Hadronic intercalibration weights for 20 GeV pions with 0, 4, 8, 9, 12, and 16 layers of the TCMT added to the calorimeter system for the configurations listed.}
\begin{tabular}{|c|c|c|c|c|c|c|c|}
\hline 
Config. & ECAL1 & ECAL2 & ECAL3 & AHCAL1 & AHCAL2 & TCMT1 & TCMT2\\
\hline
 0 & 0.0087 & 0.0092 & 0.0135 & 0.0341 & 0.0871 & 0.0000 & 0.0000 \\
 4 & 0.0085 & 0.0092 & 0.0133 & 0.0338 & 0.0643 & 0.0771 & 0.0000 \\
 8 & 0.0085 & 0.0092 & 0.0133 & 0.0336 & 0.0638 & 0.0531 & 0.0000 \\
 9 & 0.0085 & 0.0092 & 0.0133 & 0.0335 & 0.0641 & 0.0453 & 0.0926 \\
12 & 0.0084 & 0.0092 & 0.0133 & 0.0333 & 0.0635 & 0.0396 & 0.0968 \\
16 & 0.0084 & 0.0091 & 0.0132 & 0.0332 & 0.0632 & 0.0387 & 0.0940 \\
\hline
\end{tabular}
\label{tab:TableII}
\end{table}

\begin{figure}[tbh]
\centering
\includegraphics[angle=90, width=110mm]{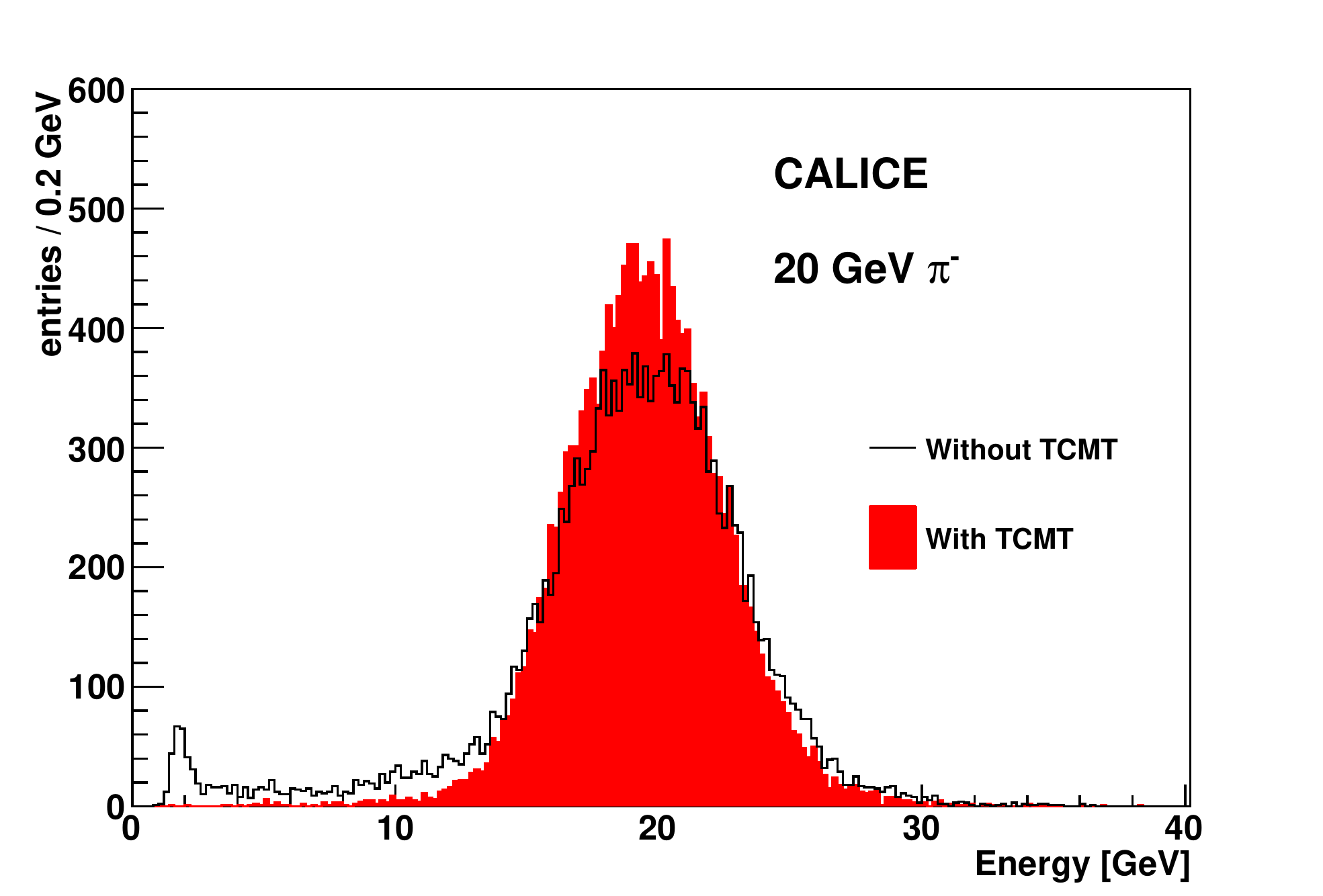}
\caption{Energy spectrum for 20 GeV pions without the TCMT (open histogram, RMS=4.7 GeV) and with the full TCMT (filled histogram, RMS=3.0 GeV). The energy from the ECAL and partial AHCAL is included.}
\label{fig:pionspectrumwithTCMT}
\end{figure}

\begin{figure}[tbh]
\centering
\includegraphics[angle=90,width=110mm]{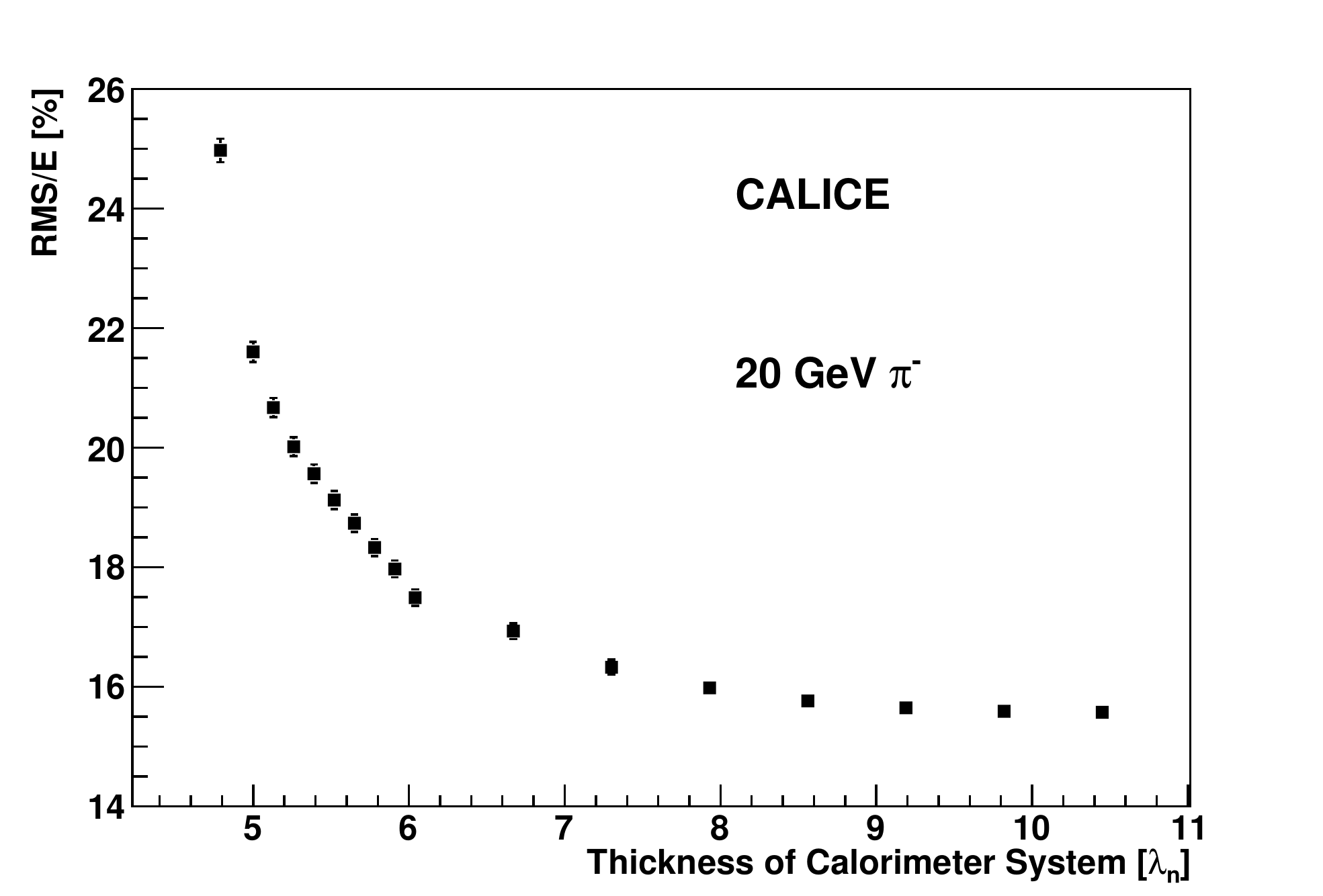}
\caption{The RMS resolution for 20 GeV pions as a function of additional TCMT layers (in units of interaction lengths). The leftmost point is the resolution for the ECAL and AHCAL with no added TCMT layers.  The rightmost point includes all 16 TCMT layers.  The calculation includes the energy from the ECAL and partial AHCAL.}
\label{fig:pionresolution}
\end{figure}

Figure \ref{fig:pionspectrumwithTCMT} compares a 20 GeV negative pion energy spectrum of the CALICE detector without the TCMT (using only the ECAL plus the partial AHCAL) with the spectrum measured adding all 16 layers of the TCMT.  The spectrum has the full set of background vetoes, was corrected for saturation, and was calculated with weights determined by the least squares minimization. The low energy tail, consisting of events not fully contained in the ECAL and AHCAL, was significantly reduced with the addition of the full TCMT.   Figure \ref{fig:pionresolution} shows the relative improvement of energy resolution as a function of the number of interaction lengths for 20 GeV pions.  To ensure sensitivity to the low tail of the spectra, especially with thinner calorimetric configurations, the resolution was calculated as the ratio of the root-mean-squared (RMS) to the mean energy.  The first point has no additional TCMT layers while the final point incorporates all 16 layers. With the full TCMT, the resolution improves 37\% relative to the partial AHCAL.   

An alternative measure of the impact of the TCMT is the hadronic shower energy leakage with and without the TCMT.  For this measure, the calorimeter system is defined as the ECAL plus the AHCAL and the first seven layers of the TCMT which corresponds to a total thickness of 5.9 nuclear interaction lengths\footnote{Similar to the thickness of a typical  ILC detector design.}.  The remaining nine layers of the TCMT are used to calculate the leakage.   
Table \ref{tab:Leakage} shows the mean leakage, RMS leakage, and fraction of events with leakage more that 10\% for 10 and 80 GeV pions.
The fractional leakage of an event is calculated as the difference between the total measured energy and the energy in the ECAL plus the AHCAL plus the seven initial TCMT layers divided by the total measured energy.   

\begin{table}[tbp]
\centering
\caption{Leakage for 10 and 80 GeV pions.}
\begin{tabular}{|c|c|c|c|}
\hline 
Energy  & Fractional & Fractional & Fraction of Events  \\
(GeV)    & Mean Leakage & RMS Leakage & Exceeding 10\% Leakage \\
\hline
 10 & $0.032 \pm 0.002$ & $0.082 \pm 0.001$ & $0.030 \pm 0.001$  \\
 80 & $0.076 \pm 0.001$ & $0.117 \pm 0.001$ & $0.061 \pm 0.001$  \\
\hline
\end{tabular}
\label{tab:Leakage}
\end{table}

Both the resolution improvement and the reduced leakage indicate that the TCMT improves the overall calorimetry and that calorimeters with seven or more interaction lengths are required to avoid leakage.  As discussed in the next section, by providing the option of calorimetric sampling after a magnetic coil, the extended depth also offers the option of thinner calorimetry inside the magnet.

\section{Magnetic Coil Emulations}
\label{sec:Section6}

The October 2006 data were used to study the impact on energy resolution of deepened calorimetry and calorimetric sampling after a magnet coil.  Most ILC detector designs include a calorimeter inside a magnet coil with a depth of about five interaction lengths. By adding layers of a combined calorimeter and muon tracker to the relatively thin hadronic calorimeter, showers could be contained and the energy resolution improved. A final ILC configuration might include a magnet coil of approximately two interaction lengths thick between the calorimeter and additional calorimetry \cite{LOIs}. 

No coil was included in the CALICE prototype but the material of a coil can be emulated with forward absorber layers of the TCMT simply by excluding the associated layers from analysis.  The emulated coil can also be moved to positions of lesser or greater interaction lengths.  Sampling fractions for these coil studies were calculated using the least squares minimization technique described earlier. The AHCAL is taken to have a depth of 3.5 interaction lengths. 
As the last two absorber layers of the AHCAL were not instrumented during the October run they are treated as leading absorber layers for the additional layers.  

\begin{figure}[tbh]
\centering
\includegraphics[width=110mm]{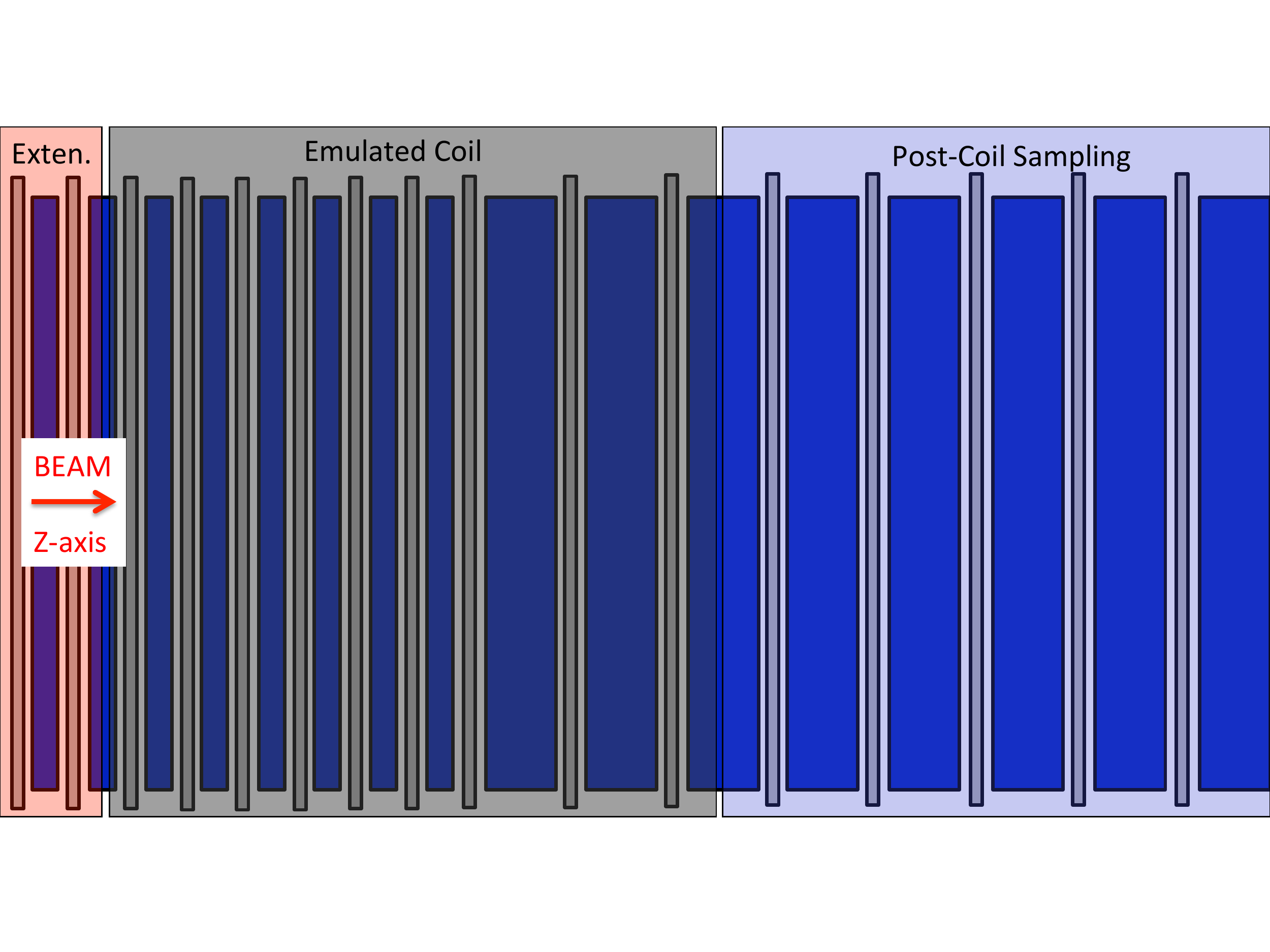}
\caption{Example coil emulation (configuration two).  The shaded rectangle on the left highlights the TCMT layers added to the AHCAL.  The central shaded rectangle highlights the portion of the TCMT used to emulate a magnet coil.  The shaded rectangle on the right highlights the TCMT layers used for post-coil sampling.  See text for additional details.}
\label{fig:emulation}
\end{figure}

Table \ref{tab:TableIII} lists the configurations of the TCMT layers used in the coil emulations. The second column indicates the number of TCMT layers (and the associated total calorimetric interaction length) added to the ECAL and AHCAL to extend the calorimeter upstream of the emulated coil. In terms of the number of TCMT layers added, the second column also identifies the first layer of the upstream edge of the emulated magnet coil. The third column gives the thickness of the emulated coil.  The configurations were chosen to ensure an emulated coil thickness near 2$\lambda_n$.  The fourth column identifies the first layer of the TCMT after the emulated coil. The final column indicates the total number of TCMT layers (and the associated interaction length) remaining to emulate the post-coil calorimetry. Figure \ref{fig:emulation} depicts configuration two, where the front two layers (highlighted in red) are treated as additional layers of the upstream calorimetry, the next nine layers (shaded in gray) are excluded from the analysis to emulate a 2.17$\lambda_n$ coil, and the final five layers of the TCMT (in blue) represent the post coil sampling.

\begin{table}[tbp]
\centering
\caption{TCMT layers used for coil emulation.}
\begin{tabular}
{|c|c|c|c|c|}
\hline
Configuration & Number of                 &  Coil Thickness  & First TCMT      & Number of  \\
Number         & TCMT layers              &   ($\lambda_n$)    & Post-Coil         & TCMT Layers  \\
                    & Pre-Coil ($\lambda_n$)  &                         & Layer Number & Post-Coil ($\lambda_n$)\\
\hline
 0 &    0 (4.8) & 1.9 & 10 & 6 (3.8)\\
 1 &    1 (5.0) & 1.7 & 10 & 6 (3.8)\\
 2 &    2 (5.1) & 2.2 & 11 & 5 (3.2)\\
 3 &    3 (5.3) & 2.0 & 11 & 5 (3.2)\\
 4 &    4 (5.4) & 1.9 & 11 & 5 (3.2)\\
 5 &    5 (5.5) & 1.8 & 11 & 5 (3.2)\\
 6 &    6 (5.6) & 1.6 & 11 & 5 (3.2)\\
 7 &    7 (5.8) & 2.2 & 12 & 4 (2.5)\\
 8 &    8 (5.9) & 2.0 & 12 & 4 (2.5)\\
 9 &    9 (6.0) & 1.9 & 12 & 4 (2.5)\\
10 & 10 (6.7) & 1.9 & 13 & 3 (1.9)\\
11 & 11 (7.3) & 1.9 & 14 & 2 (1.3)\\
12 & 12 (7.9) & 1.9 & 15 & 1 (0.6)\\
\hline
\end{tabular}
\label{tab:TableIII}
\end{table}

\begin{figure}[tbh]
\centering
\includegraphics[angle=90, width=110mm]{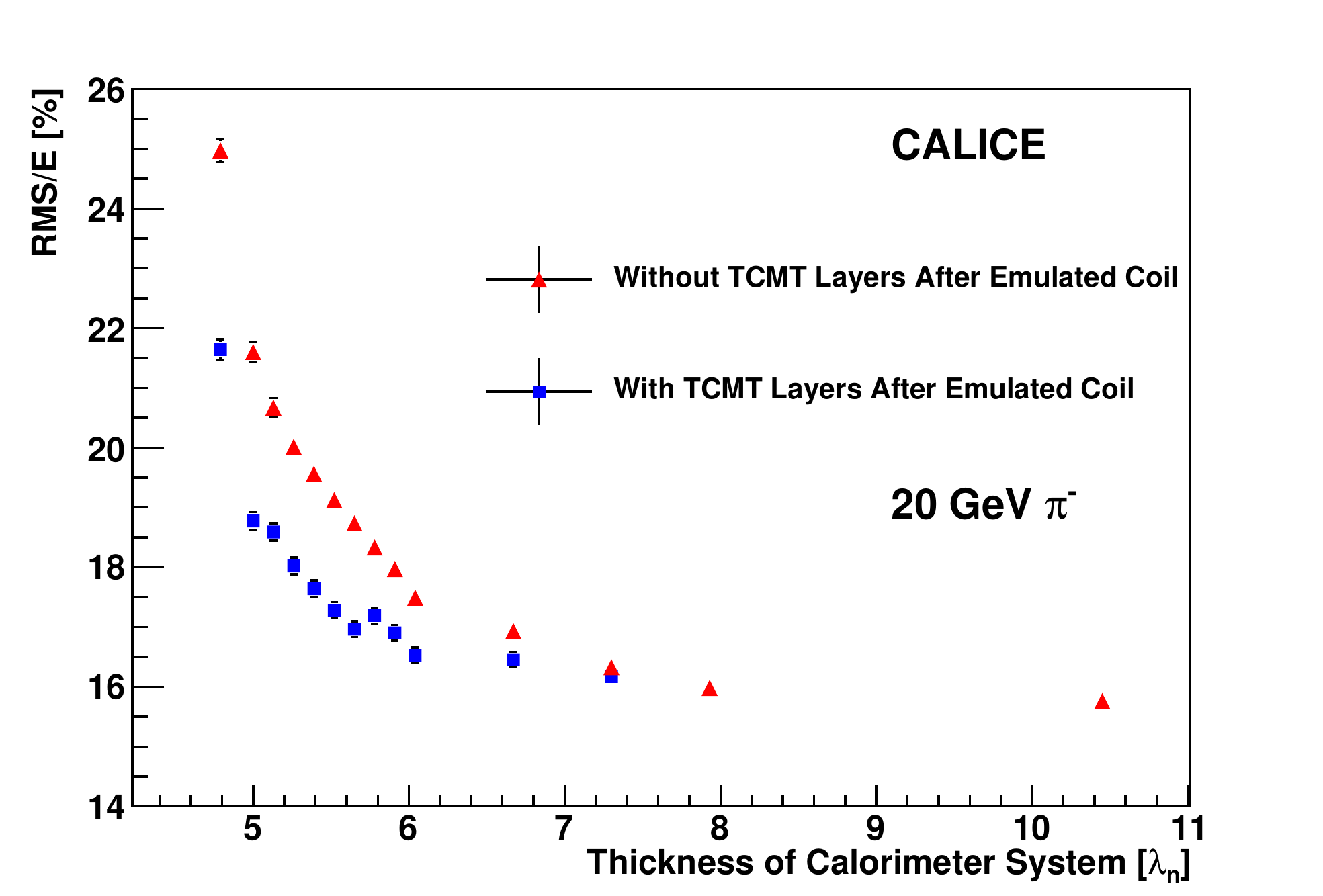}
\caption{Comparison of the energy RMS resolution of a 20 GeV negative pion sample with an emulated coil without final TCMT layers after the coil (triangular symbols) and with final TCMT layers after the coil (square symbols). The calculation includes the energy from the ECAL and partial AHCAL. }
\label{fig:coilsimulation}
\end{figure}

Figure  \ref{fig:coilsimulation} shows the impact of post-coil sampling as a function of the thickness of the main calorimeter, by comparing the energy resolution with and without TCMT layers behind an emulated magnetic coil. The red triangular symbols show the energy resolution of a 20 GeV negative pion beam for a calorimeter system incorporating the ECAL, the partial AHCAL, and additional layers of the TCMT (here the abscissa corresponds to the full depth of the calorimetry). The energy resolution is calculated as the root-mean-square of the energy distribution divided by the average energy. These are the points shown in Fig. \ref{fig:pionresolution}.  The leftmost triangular symbol corresponds to no calorimeter extension and the rightmost to extension with the full TCMT. 

The blue square symbols in Fig. \ref{fig:coilsimulation} show the energy resolution for a system including the ECAL, the partial AHCAL, forward layers of the TCMT, an emulated coil, and the remaining layers of the TCMT (in this case the abscissa corresponds to the position of the forward edge of the emulated coil).  The leftmost lower square symbol corresponds to a coil immediately after the HCAL followed by additional sampling with the remaining ten TCMT layers (configuration 0 in Table \ref{tab:TableIII}). Relative to a calorimeter extension only, the post-coil sampling improves resolution from 25\% to 21.5\%.  At a depth of 5.5$\lambda_n$, the additional sampling layers improve energy resolution for a 20 GeV pion by a relative 8\%. The rightmost points in Fig. \ref{fig:coilsimulation} correspond to an HCAL extended by the full length of the TCMT or a coil with little or no additional sampling.

\begin{figure}[tbh]
\centering
\includegraphics[angle=90, width=110mm]{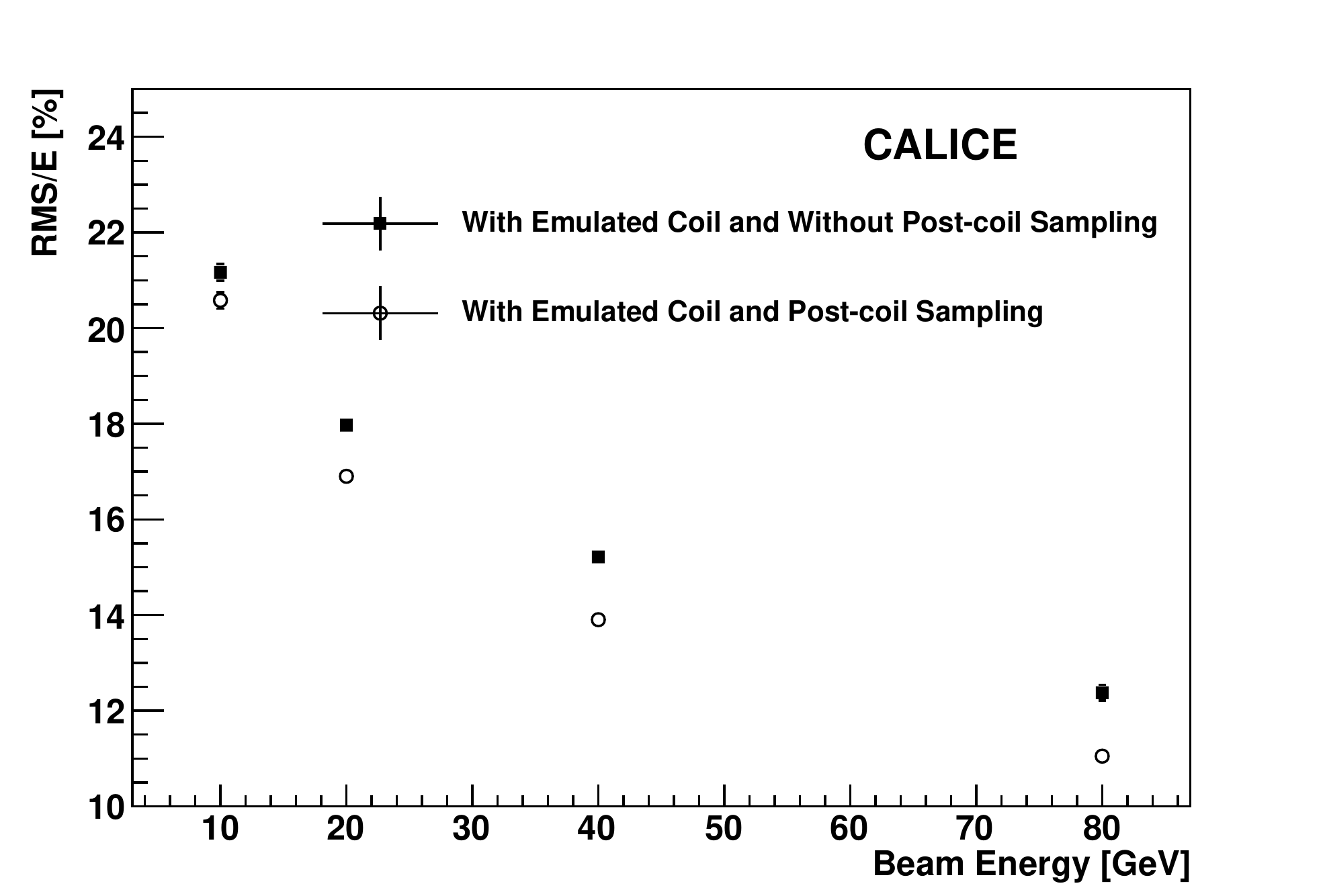}
\caption{ Energy RMS resolution versus beam energy for at 5.5 $\lambda_n$ calorimeter system without a coil (solid squares) and followed by an emulated coil and post coil sampling (open circles). The  calculation includes the energy from the ECAL and partial AHCAL.}
\label{fig:coilenergy}
\end{figure}

Figure \ref{fig:coilenergy} indicates that for a coil located at 5.5$\lambda_n$ the improvement due to post-coil sampling increases with energy.  For this configuration the relative resolution improvement at 20 GeV is 6\% and at 80 GeV 16\%.   The sampling weights derived for 20 GeV were used at all energies.

\section{Summary and Conclusions}
\label{sec:Section7}

Prototypes for a high granularity calorimeter designed to use particle flow algorithms have been developed by CALICE in order to achieve the excellent energy resolution required for lepton collider detectors. A limitation on detector resolution is the depth of the hadron calorimeter contained within a magnet coil. The hadronic calorimeter thickness in ILC detector designs is usually about 4-5 nuclear interaction lengths which is too thin to contain the most energetic showers and late showering pions. To compensate for this deficiency, the TCMT was designed to extend calorimetry and, in addition, function as a muon tracking system. The prototype design also provided an opportunity to study the use of a number of SiPMs in particle detectors and to emulate the impact on the energy resolution of a magnet coil in various positions.

The extruded scintillator/SiPM TCMT performed well under beam conditions.  The detector was robust, stable and efficient.  The mean light yield per muon was 5.8 photoelectrons.  However, attenuation of about 10 to 20\% per meter was evident.  The TCMT contains hadronic shower energy leakage and improves energy resolution. Inclusion of all sixteen layers of the TCMT to the 5.0$\lambda_n$ CALICE calorimeter in the October 2006 test beam configuration (which included the ECAL and a partial AHCAL) improves pion resolution by 9.3\% at 20 GeV, a relative improvement of 37\%.   For a coil situated outside a 5.5$\lambda_{n}$ calorimeter system, a typical configuration for proposed ILC detectors, a TCMT would improve relative energy resolution by 6-16\% between 20 and 80 GeV.  Furthermore, additional sampling will reduce extreme fluctuations in energy measurements which will help minimize background to processes characterized by missing transverse energy.  For a coil beyond 7$\lambda_n$, post-coil sampling provides negligible improvement for a 20 GeV pion.  In all cases the resolution was taken as the RMS to ensure sensitivity to distribution tails or energy leakage.

\acknowledgments
We would like to thank the technicians and the engineers who
contributed to the design and construction of the TCMT, in
particular P. Stone. FNAL also made essential contributions to the 
construction of the TCMT mechanical structure.
We also
gratefully acknowledge the DESY, CERN, and FNAL management for their support and
hospitality, and their accelerator staff for the reliable and efficient
beam operation. 
This work was supported by the 
Bundesministerium f\"{u}r Bildung und Forschung, Germany;
by the  the DFG cluster of excellence `Origin and Structure of the Universe' of Germany ; 
by the Helmholtz-Nachwuchsgruppen grant VH-NG-206;
by the BMBF, grant no. 05HS6VH1;
by the Alexander von Humboldt Foundation (Research Award IV, RUS1066839 GSA);
by joint Helmholtz Foundation and RFBR grant HRJRG-002, SC Rosatom;
by Russian Grants  SS-1329.2008.2 and RFBR08-02-121000-0FI
and by the Russian Ministry of Education and Science contract 02.740.11.0239;
by MICINN and CPAN, Spain;
by CRI(MST) of MOST/KOSEF in Korea;
by the US Department of Energy and the US National Science
Foundation;
by the Ministry of Education, Youth and Sports of the Czech Republic
under the projects AV0 Z3407391, AV0 Z10100502, LC527  and LA09042  and by the
Grant Agency of the Czech Republic under the project 202/05/0653;  
and by the Science and Technology Facilities Council, UK.

\end{document}

%% file: calice_authors_tcmt_04jan2012.tex
\author{\centering 
\LARGE\bf The CALICE Collaboration}

\author{
C.\,Adloff, 
J.\,Blaha, 
J.-J.\,Blaising, 
C.\,Drancourt,
A.\,Espargili\`{e}re, 
R.\,Gaglione, 
N.\,Geffroy, 
Y.\,Karyotakis, 
J.\,Prast,
G.\,Vouters
 \\
Laboratoire d'Annecy-le-Vieux de Physique des Particules, Universit\'{e} de Savoie,
CNRS/IN2P3,
9 Chemin de Bellevue BP110, F-74941 Annecy-le-Vieux CEDEX, France
}

\author{
B.\,Bilki$^a$, 
K.\,Francis,
J.\,Repond, 
J.\,Smith$^b$, 
L.\,Xia 
\\
Argonne National Laboratory,
9700 S.\ Cass Avenue,
Argonne, IL 60439-4815,
USA
}

\author{
E.\,Baldolemar, 
J.\,Li$^c$, 
S.\,T.\,Park, 
M.\,Sosebee, 
A.\,P.\,White, 
J.\,Yu
 \\
Department of Physics, SH108, University of Texas, Arlington, TX 76019, USA
}

\author{
T.\,Buanes, G.\,Eigen
 \\
University of Bergen, Inst.\, of Physics, Allegaten 55, N-5007 Bergen, Norway
}

\author{
Y.\,Mikami, 
N.\,K.\,Watson 
\\
University of Birmingham,
School of Physics and Astronomy,
Edgbaston, Birmingham B15 2TT, UK
}

\author{
G.\,Mavromanolakis$^d$, 
M.\,A.\,Thomson, 
D.\,R.\,Ward, 
W.\,Yan$^e$, 
\\
University of Cambridge, Cavendish Laboratory, J J Thomson Avenue, CB3 0HE, UK
}

\author{
D.\,Benchekroun, 
A.\,Hoummada, 
Y.\,Khoulaki
\\
Universit\'{e} Hassan II A\"{\i}n Chock, Facult\'{e} des sciences.\, B.P. 5366 Maarif, Casablanca, Morocco
}

\author{
M.\,Benyamna, 
C.\,C\^{a}rloganu, 
F.\,Fehr, 
P.\,Gay, 
S.\,Manen, 
L.\,Royer
\\
Clermont Univertsit\'e, Universit\'e Blaise Pascal, CNRS/IN2P3, LPC, BP
10448, F-63000 Clermont-Ferrand, France
}

\author{
G.\,C.\,Blazey$^*$,
S.\,Boona,
D.\,Chakraborty,
A.\,Dyshkant, 
D.\,Hedin,
J.\,G.\,R.\,Lima,
J.\,Powell, 
V.\,Rykalin,
N.\,Scurti,
M.\,Smith,
N.\,Tran,
V.\,Zutshi
\\
NICADD, Northern  Illinois University,
Department of Physics,
DeKalb, IL 60115,
USA
}

\author{
J.\,-Y.\,Hostachy, 
L.\,Morin
\\
Laboratoire de Physique Subatomique et de Cosmologie - Universit\'{e} Joseph Fourier Grenoble 1 -
CNRS/IN2P3 - Institut Polytechnique de Grenoble,
53, rue des Martyrs,
38026 Grenoble CEDEX, France
}

\author{
U.\,Cornett, 
D.\,David, 
J.\,Dietrich, 
G.\,Falley, 
K.\,Gadow, 
P.\,G\"{o}ttlicher, 
C.\,G\"{u}nter,
B.\,Hermberg, 
S.\,Karstensen, 
F.\,Krivan,
A.\,-I.\,Lucaci-Timoce$^d$, 
S.\,Lu, 
B.\,Lutz, 
I.\,Marchesini, 
S.\,Morozov, 
V.\,Morgunov$^f$, 
M.\,Reinecke, 
F.\,Sefkow, 
P.\,Smirnov,
M.\,Terwort,
A.\,Vargas-Trevino, 
\\
DESY, Notkestrasse 85,
D-22603 Hamburg, Germany
}

\author{
N.\,Feege, 
E.\,Garutti
\\
Univ. Hamburg,
Physics Department,
Institut f\"ur Experimentalphysik,
Luruper Chaussee 149,
22761 Hamburg, Germany
}

\author{
P.\,Eckert,
A.\,Kaplan,
 H.\,-Ch.\,Schultz-Coulon,
 W.\,Shen,
 R.\,Stamen,
 A.\,Tadday
\\
 University of Heidelberg, Fakultat fur Physik und Astronomie,
Albert Uberle Str. 3-5 2.OG Ost,
D-69120 Heidelberg, Germany
}

\author{
 E.\,Norbeck, 
Y.\,Onel
\\
University of Iowa, Dept. of Physics and Astronomy,
203 Van Allen Hall, Iowa City, IA 52242-1479, USA
}

\author{
G.\,W.\,Wilson
\\
University of Kansas, Department of Physics and Astronomy,
Malott Hall, 1251 Wescoe Hall Drive, Lawrence, KS 66045-7582, USA
}

\author{
K.\,Kawagoe,  
S.\,Uozumi$^g$
\\
 Department of Physics, Kobe University, Kobe, 657-8501, Japan
}

\author{
P.\,D.\,Dauncey, 
A.\,-M.\,Magnan
\\
Imperial College London, Blackett Laboratory,
Department of Physics,
Prince Consort Road,
London SW7 2AZ, UK 
}

\author{
V.\,Bartsch$^h$, 
M.\,Wing
\\
Department of Physics and Astronomy, University College London,
Gower Street,
London WC1E 6BT, UK
}

\author{
F.\,Salvatore$^h$ 
\\
Royal Holloway University of London,
Dept. of Physics,
Egham, Surrey TW20 0EX, UK
}

\author{
E.\,Calvo~Alamillo, 
M.-C.\, Fouz, 
J.\,Puerta-Pelayo 
\\
CIEMAT, Centro de Investigaciones Energeticas, Medioambientales y Tecnologicas, Madrid, Spain 
}

\author{
B.\,Bobchenko, 
M.\,Chadeeva, 
M.\,Danilov, 
A.\,Epifantsev, 
O.\,Markin,
R.\,Mizuk, 
E.\,Novikov, 
V.\,Rusinov, 
E.\,Tarkovsky
\\
Institute of Theoretical and Experimental Physics, B. Cheremushkinskaya ul. 25,
RU-117218 Moscow, Russia
}

\author{
N.\,Kirikova,
V.\,Kozlov, 
Y.\,Soloviev 
\\
P.\,N.\, Lebedev Physical Institute,
Russian Academy of Sciences,
117924 GSP-1 Moscow, B-333, Russia
}

\author{
P.\,Buzhan, B.\,Dolgoshein$^c$, 
A.\,Ilyin, V.\,Kantserov, V.\,Kaplin, A.\,Karakash, E.\,Popova, S.\,Smirnov 
\\
Moscow Physical Engineering Inst., MEPhI,
Dept. of Physics,
31, Kashirskoye shosse,
115409 Moscow, Russia
}

\author{
A.\,Frey$^i$, 
C.\,Kiesling,
K.\,Seidel, 
F.\,Simon,
C.\,Soldner, 
L.\,Weuste
\\
Max Planck Inst. f\"ur Physik,
F\"ohringer Ring 6,
D-80805 Munich, Germany
}

\author{
J.\,Bonis, 
B.\,Bouquet,    
S.\,Callier, 
P.\,Cornebise, 
Ph.\,Doublet,
F.\,Dulucq, 
M.\,Faucci Giannelli, 
J.\,Fleury,
H.\,Li$^j$, 
G.\,Martin-Chassard, 
F.\,Richard, 
Ch.\,de la Taille, 
R.\,P\"{o}schl, 
L.\,Raux,  
N.\,Seguin-Moreau, 
F.\,Wicek
\\
Laboratoire de l'Acc\'{e}l\'{e}rateur Lin\'{e}aire, Centre
Scientifique d'Orsay, Universit\'{e} de Paris-Sud XI, CNRS/IN2P3, BP
34, B\^atiment 200, F-91898 Orsay CEDEX, France
}

\author{
M.\,Anduze,
V.\,Boudry, 
J-C.\,Brient, 
D.\,Jeans, 
P.\,Mora de Freitas, 
G.\,Musat, 
M.\,Reinhard, 
M.\,Ruan,  
H.\,Videau
\\
 Laboratoire Leprince-Ringuet (LLR)  -- \'{E}cole Polytechnique, CNRS/IN2P3, F-91128 Palaiseau, France
}

\author{
B.\,Bulanek,
J.\,Zacek 
\\
Charles University, Institute of Particle \& Nuclear Physics,
V Holesovickach 2,
CZ-18000 Prague 8, Czech Republic  
}

\author{
J.\,Cvach, 
P.\,Gallus, 
M.\,Havranek, 
M.\,Janata, 
J.\,Kvasnicka,
D.\,Lednicky,
M.\,Marcisovsky, 
I.\,Polak, 
J.\,Popule, 
L.\,Tomasek, 
M.\,Tomasek, 
P.\,Ruzicka, 
P.\,Sicho, 
J.\,Smolik, 
V.\,Vrba, 
J.\,Zalesak 
\\
Institute of Physics, Academy of Sciences of the Czech Republic, Na Slovance 2,
CZ-18221 Prague 8, Czech Republic
}

\author{
B.\,Belhorma,
H.\,Ghazlane
\\
Centre National de l'Energie, des Sciences et des Techniques Nucl\'{e}aires, 
B.P. 1382, R.P. 10001, Rabat, Morocco
}

\author{
T.\,Takeshita
\\
Shinshu Univ.\,,
Dept. of Physics,
3-1-1 Asaki,
Matsumoto-shi, Nagano 390-861,
Japan
}

\author{ \\
\llap{$^*$}Corresponding author: \email{gblazey@nicadd.niu.edu}\\
\llap{$^a$}Also at University of Iowa\\
\llap{$^b$}Also at University of Texas, Arlington\\
\llap{$^c$}Deceased\\
\llap{$^d$}Now at CERN\\
\llap{$^e$}Now at Dept.\, of Modern Physics, Univ. of Science and Technology of China, 96 Jinzhai Road, Hefei, Anhui, 230026, P.\, R.\, China\\
\llap{$^f$}On leave from ITEP\\
\llap{$^g$}Now at Kyungpook National University, South Korea\\
\llap{$^h$}Now at University of Sussex, Physics and
Astronomy Department, Brighton, Sussex, BN1 9QH, UK\\
\llap{$^i$}Now at University of G\"{o}ttingen, II. Physikalisches Institut
Friedrich-Hund-Platz 1, 37077 G\"ottingen, Germany\\
\llap{$^j$}Now at LPSC Grenoble\\

}

\newpage
